\documentclass[twocolumn]{aastex63}

\shorttitle{Triply imaged red compact object at $z_{\mathrm{phot}}\simeq7.6$}
\shortauthors{L. J. Furtak et al.}

\usepackage[encapsulated]{CJK}
\usepackage{amsmath}

\tabletypesize{\footnotesize}
\begin{document}

%%%%%%%%%%%%%%%%%%%%%%%%%%%%%%%%%%%%%%%%%%%%%%%%%%%%
\title{JWST UNCOVER: Extremely red and compact object at $z_{\mathrm{phot}}\simeq7.6$ triply imaged by Abell 2744}

%%%%%%%%%%%%%%%%%%%%%%%%%%%%%%%%%%%%%%%%%%%%%%%%%%%%

\correspondingauthor{Lukas J. Furtak}
\email{furtak@post.bgu.ac.il}

\author[0000-0001-6278-032X]{Lukas J. Furtak}
\affiliation{Physics Department, Ben-Gurion University of the Negev, P.O. Box 653, Be'er-Sheva 84105, Israel}

\author[0000-0002-0350-4488]{Adi Zitrin}
\affiliation{Physics Department, Ben-Gurion University of the Negev, P.O. Box 653, Be'er-Sheva 84105, Israel}

\author[0000-0003-0390-0656]{Ad\`{e}le Plat}
\affiliation{Steward Observatory, University of Arizona, 933 N Cherry Ave, Tucson, AZ 85721 USA}

\author[0000-0001-7201-5066]{Seiji Fujimoto}
\altaffiliation{Hubble Fellow}
\affiliation{Department of Astronomy, The University of Texas at Austin, Austin, TX 78712, USA}

\author[0000-0001-9269-5046]{Bingjie Wang (\begin{CJK*}{UTF8}{gbsn}王冰洁\ignorespacesafterend\end{CJK*})}
\affiliation{Department of Astronomy \& Astrophysics, The Pennsylvania State University, University Park, PA 16802, USA}
\affiliation{Institute for Computational \& Data Sciences, The Pennsylvania State University, University Park, PA 16802, USA}
\affiliation{Institute for Gravitation and the Cosmos, The Pennsylvania State University, University Park, PA 16802, USA}

\author[0000-0002-7524-374X]{Erica J. Nelson}
\affiliation{Department for Astrophysical and Planetary Science, University of Colorado, Boulder, CO 80309, USA}

\author[0000-0002-2057-5376]{Ivo Labb{\'e}}
\affiliation{Centre for Astrophysics and Supercomputing, Swinburne University of Technology, Melbourne, VIC 3122, Australia}

\author[0000-0001-5063-8254]{Rachel Bezanson}
\affiliation{Department of Physics and Astronomy and PITT PACC, University of Pittsburgh, Pittsburgh, PA 15260, USA}

\author[0000-0003-2680-005X]{Gabriel B. Brammer}
\affiliation{Cosmic Dawn Center (DAWN), Niels Bohr Institute, University of Copenhagen, Jagtvej 128, K{\o}benhavn N, DK-2200, Denmark}

\author[0000-0002-8282-9888]{Pieter van Dokkum}
\affiliation{Department of Astronomy, Yale University, New Haven, CT 06511, USA}

\author[0000-0003-4564-2771]{Ryan Endsley}
\affiliation{Department of Astronomy, The University of Texas at Austin, Austin, TX 78712, USA}

\author[0000-0002-3254-9044]{Karl Glazebrook}
\affiliation{Centre for Astrophysics and Supercomputing, Swinburne University of Technology, Melbourne, VIC 3122, Australia} 

\author[0000-0002-5612-3427]{Jenny E. Greene}
\affiliation{Department of Astrophysical Sciences, 4 Ivy Lane, Princeton University, Princeton, NJ 08544, USA}

\author[0000-0001-6755-1315]{Joel Leja}
\affiliation{Department of Astronomy \& Astrophysics, The Pennsylvania State University, University Park, PA 16802, USA}
\affiliation{Institute for Computational \& Data Sciences, The Pennsylvania State University, University Park, PA 16802, USA}
\affiliation{Institute for Gravitation and the Cosmos, The Pennsylvania State University, University Park, PA 16802, USA}

\author[0000-0002-0108-4176]{Sedona H. Price}
\affiliation{Department of Physics and Astronomy and PITT PACC, University of Pittsburgh, Pittsburgh, PA 15260, USA}

\author[0000-0001-8034-7802]{Renske Smit}
\affiliation{Astrophysics Research Institute, Liverpool John Moores University, 146 Brownlow Hill, Liverpool L3 5RF, UK}

\author[0000-0001-6106-5172]{Daniel P. Stark}
\affiliation{Steward Observatory, University of Arizona, 933 N Cherry Ave, Tucson, AZ 85721 USA}

\author[0000-0003-1614-196X]{John R. Weaver}
\affiliation{Department of Astronomy, University of Massachusetts, Amherst, MA 01003, USA}

\author[0000-0001-7160-3632]{Katherine E. Whitaker}
\affiliation{Department of Astronomy, University of Massachusetts, Amherst, MA 01003, USA}
\affiliation{Cosmic Dawn Center (DAWN), Niels Bohr Institute, University of Copenhagen, Jagtvej 128, K{\o}benhavn N, DK-2200, Denmark} 

\author[0000-0002-7570-0824]{Hakim Atek}
\affiliation{Institut d'Astrophysique de Paris, CNRS, Sorbonne Universit\'e, 98bis Boulevard Arago, 75014, Paris, France}

\author[0000-0002-7636-0534]{Jacopo Chevallard}
\affiliation{Department of Physics, University of Oxford, Denys Wilkinson Building, Keble Road, Oxford OX1 3RH, UK}

\author[0000-0002-9551-0534]{Emma Curtis-Lake}
\affiliation{Centre for Astrophysics Research, Department of Physics, Astronomy and Mathematics, University of Hertfordshire, Hatfield AL10
9AB, UK}

\author[0000-0001-8460-1564]{Pratika Dayal}
\affiliation{Kapteyn Astronomical Institute, University of Groningen, P.O. Box 800, 9700 AV Groningen, The Netherlands}

\author[0000-0001-6865-2871]{Anna Feltre}
\affiliation{INAF -- Osservatorio di Astrofisica e Scienza dello Spazio, Via Gobetti 93/3, 40129, Bologna, Italy}

\author[0000-0002-8871-3026]{Marijn Franx}
\affiliation{Leiden Observatory, Leiden University, P.O.Box 9513, NL-2300 AA Leiden, The Netherlands}

\author[0000-0001-7440-8832]{Yoshinobu Fudamoto}
\affiliation{Waseda Research Institute for Science and Engineering, Faculty of Science and Engineering, Waseda University, 3-4-1 Okubo, Shinjuku, Tokyo 169-8555, Japan}
\affiliation{National Astronomical Observatory of Japan, 2-21-1, Osawa, Mitaka, Tokyo, Japan}

\author[0000-0001-9002-3502]{Danilo Marchesini}
\affiliation{Department of Physics and Astronomy, Tufts University, 574 Boston Ave., Medford, MA 02155, USA}

\author[0000-0002-8530-9765]{Lamiya A. Mowla}
\affiliation{Dunlap Institute for Astronomy and Astrophysics, 50 St. George Street, Toronto, Ontario, M5S 3H4, Canada}

\author[0000-0002-9651-5716]{Richard Pan}
\affiliation{Department of Physics and Astronomy, Tufts University, 574 Boston Ave., Medford, MA 02155, USA}

\author[0000-0002-1714-1905]{Katherine A. Suess}
\affiliation{Department of Astronomy and Astrophysics, University of California, Santa Cruz, 1156 High Street, Santa Cruz, CA 95064 USA}
\affiliation{Kavli Institute for Particle Astrophysics and Cosmology and Department of Physics, Stanford University, Stanford, CA 94305, USA}

\author[0000-0001-6798-9202]{Alba Vidal-Garc\'{i}a}
\affiliation{Observatorio Astron\'{o}mico Nacional, C/ Alfonso XII 3, 28014 Madrid, Spain}

\author[0000-0003-2919-7495]{Christina C. Williams}
\affiliation{NSF’s National Optical-Infrared Astronomy Research Laboratory, 950 N. Cherry Avenue, Tucson, AZ 85719, USA}
\affiliation{Steward Observatory, University of Arizona, 933 N Cherry Ave, Tucson, AZ 85721 USA}

%%%%%%%%%%%%%%%%%%%%%%%%%%%%%%%%%%%%%%%%%%%%%%%%%%%%

\begin{abstract}
Recent JWST/NIRCam imaging taken for the ultra-deep UNCOVER program reveals a very red dropout object at  $z_{\mathrm{phot}}\simeq7.6$, triply imaged by the galaxy cluster Abell~2744 ($z_{\mathrm{d}}=0.308$). All three images are very compact, i.e. unresolved, with a de-lensed size upper-limit of $r_{e}\lesssim35$\,pc. The images have apparent magnitudes of $m_{\mathrm{F444W}}\sim25-26$\,AB, and the magnification-corrected absolute UV magnitude of the source is $M_{\mathrm{UV},1450}=-16.81\pm0.09$. From the sum of observed fluxes and from a spectral energy distribution (SED) analysis, we obtain estimates of the bolometric luminosities of the source of $L_{\mathrm{bol}}\gtrsim10^{43}\,\frac{\mathrm{erg}}{\mathrm{s}}$ and $L_{\mathrm{bol}}\sim10^{44}-10^{46}\,\frac{\mathrm{erg}}{\mathrm{s}}$, respectively. Based on its compact, point-like appearance, its position in color-color space and the SED analysis, we tentatively conclude that this object is a UV-faint dust-obscured quasar-like object, i.e. an active galactic nucleus (AGN) at high redshift. We also discuss other alternative origins for the object's emission features, including a massive star cluster, Population~III, supermassive, or dark stars, or a direct-collapse black hole. Although populations of red galaxies at similar photometric redshifts have been detected with JWST, this object is unique in that its high-redshift nature is corroborated geometrically by lensing, that it is unresolved despite being magnified -- and thus intrinsically even more compact -- and that it occupies notably distinct regions in both size-luminosity and color-color space. Planned UNCOVER JWST/NIRSpec observations, scheduled in Cycle~1, will enable a more detailed analysis of this object.
\end{abstract}

%%%%%%%%%%%%%%%%%%%%%%%%%%%%%%%%%%%%%%%%%%%%%%%%%%%%

%%%%%%%%%%%%%%%%%%%%%%%%%%%%%%%%%%%%%%%%%%%%%%%%%%%%
\section{Introduction}
\label{sec:intro}
Quasars, or \emph{quasi-stellar objects}, are extremely luminous objects powered by supermassive black holes (SMBH) in their centers. Accretion onto the SMBH transfers a large amount of potential energy from the in-falling matter, and kinetic energy due to friction, into thermal energy, which in turn results in a very high luminosity (typical bolometric luminosities of $L_{\mathrm{bol}}\sim10^{44}-10^{48}\,\frac{\mathrm{erg}}{\mathrm{s}}$; see e.g. \citealt{Shen2020QuasarLF}).

While quasars are known in relatively large numbers throughout the Universe and especially out to $z\sim6$ \citep[e.g.][]{Banados2016100PS1Quasars,Lyke2020,Flesch2021}, only several quasars are known at high-redshifts $z\gtrsim7$, albeit with increasing numbers (e.g. \citealt{Banados2018z7p54Quasar,Wang2018z7Quasar,Jinyi2021Quasars37z6}). Nevertheless, the formation and evolution of these high-redshift SMBHs, observed when the Universe was less than 1\,Gyr old, is poorly understood, as the accretion rate onto them, or alternatively their initial masses, seem largely prohibited by common formation scenarios \citep[e.g.][and references therein; though see also \citealt{Trakhtenbrot2017ApJ...836L...1T}]{Volonteri2012BHformation,Fan2019BAAS...51c.121F}. Moreover, while the quasar luminosity function (LF) implies a larger abundance of fainter objects \citep[i.e. faint end slopes of $\sim1.2-1.6$; e.g.][]{Glikman2011ApJ...728L..26G,Niida_2020} similar to the galaxy LF, faint quasars with $M_{\mathrm{UV}}\gtrsim-22$ seem to be rare in observations. Indeed, the JWST \citep{Gardner2006,McElwain2023}, launched one year ago, may help in finding more such objects.
Recent work suggests that active galactic nuclei (AGN) that are excessively massive relative to their host galaxies, accreting at high Eddington rates, would be detectable with JWST at high redshifts \citep[][]{Volonteri2022AGNz9} and potentially shed light on early black-hole growth.

The last decade has seen many hundreds of high-redshift objects detected with the \textit{Hubble Space Telescope} (HST) and with several ground-based surveys such as UltraVISTA \citep{Ultravista2012A&A...544A.156M} or LAGER \citep{LAGER2017ApJ...842L..22Z}, for example. It has by now become well established that the universe was reionized in the first billion years, i.e. by redshift $z\sim5.5-6$ \citep[e.g.][]{Fan2006Quasars,Stark2010z3-7fractions,Pentericci2011Lyfractionz7,Robertson15,Planck15,Bosman2022MNRAS.514...55B}. However, it is not yet clear if early galaxies supply sufficient ionizing radiation to account for reionization or if a major contribution from quasars or other exotic sources (e.g. supermassive stars, X-ray binaries, etc.) is needed. The shape of the galaxy ultra-violet (UV) LF at high redshifts implies that most of the ionizing radiation originated from the more abundant population of faint galaxies \citep[e.g.][]{atek15b}, but these have been largely beyond the reach of HST. It is also not clear if the LF already shows a turnover at HST depth, even in the deepest lensed fields \citep[e.g.][]{atek15b,Bouwens2017}, although a tentative turnover may have been detected \citep{bouwens2022b,atek18}. One of the main goals of the JWST is to study the first stars and galaxies. In combination with the power of gravitational lensing, we will indeed be able to address some of these key questions, as can already be implied by the first six months of JWST operations \citep[e.g.][]{Adams2023MNRAS.518.4755A,Bouwens2022LFJWST,Atek2023,furtak22b,Castellano2022arXiv221206666C,Donnan2022highz,Finkelstein2022arXiv221105792F,Morishita2022arXiv221109097M}.

Another crucial ingredient to understanding the epoch of cosmic reionization (EoR) is the spectrum of high-redshift galaxies. Some of the highest-redshift objects spectroscopically confirmed from the ground \citep{Oesch20157p73,Zitrin2015Lyalpha,Roberts_Borsani2015,Stark2017} seem to hint at a harder UV spectrum than expected from typical stellar populations \citep{Stark2014CIIIdetectionz67,Stark2015CIV}, indicating possible AGN activity, Population~III (Pop.~III) star contribution, or other hard ionizing photon sources \citep[e.g.][]{Mainali2017,Laporte2017,Matthee2020CR7}. Observations with the JWST have already supplied unprecedented rest-frame UV and optical spectra for some very high-redshift galaxies \citep{Roberts-Borsani2022arXiv221015639R,Williams2022z9p5,Curtis-Lake2022arXiv221204568C} and might indeed supply new insight soon. In addition, objects that potentially bridge the typical high-redshift galaxy and AGN populations were also recently reported. \citet{Fujimoto2022Natur.604..261F} found a dusty $M_{\mathrm{UV},1450}\simeq-23.2$ compact object bridging galaxies and quasars in the EoR. Its SED was distinct from typical high-redshift galaxies but could be well explained by the combination of a dusty star-forming galaxy (DSFG) SED and that of a quasar (see also \citealt{Matthee2021colors,Cui2021NatAs...5.1069C} for discussion of AGN and high-redshift galaxy colors). The \citet{Fujimoto2022Natur.604..261F} galaxy is faint in X-rays, indicating the emergence of a uniquely UV-compact star-forming region or a Compton-thick super-Eddington black hole accretion disk at the dusty starburst core. Additionally, \citet{Endsley2022AGN_arXiv220600018E} reported a heavily-obscured hyper-luminous AGN at $z=6.85$ that was originally identified as very dusty Lyman-break $z\simeq7$ galaxy with strong far-infrared (FIR) and radio emission \citep{Endsley2022AGNcandidate_MNRAS.512.4248E}. As another example, \citet{Onoue2022arXiv220907325O} have recently detected a faint ($M_{\mathrm{UV},1450}\simeq-19.5$), relatively low-mass AGN candidate at $z\sim5$ in the JWST \textit{Cosmic Evolution Early Release Science Survey} program \citep[CEERS;][]{Finkelstein2022arXiv221105792F,bagley2022}, based on its compact shape and red spectrum which can be attributed to the redshifted rest-frame optical H$\beta$ and [\ion{O}{3}] and H$\alpha$ emission lines. Since then, several high-redshift ($z\sim4-9$) AGN were discovered and spectroscopically confirmed in the CEERS field \citep[][]{Kocevski2023,Yang2023CEERSAGN,Larson2023,Harikane2023,Akins2023} as well as another possible type of red source, a quiescent galaxy at $z\sim5$ \citep{Carnall2023}.

In this work, we highlight the discovery a very compact high-redshift ($z\gtrsim7$) object that is triply imaged by the galaxy cluster Abell~2744 ($z_{\mathrm{d}}=0.308$) in recent JWST imaging. Two of the multiple images of this system were originally found by \citet{Atek2014A2744} in \textit{Hubble Frontier Fields} \citep[HFF;][]{Lotz2016HFF} data. The new JWST imaging now allows us to secure the third counter image and hint at the peculiar nature of this object: they accentuate the compact, potentially point-source like nature of the source, and reveal a spectral energy distribution (SED) with very red shape towards the longer wavelengths. The size, colors, and SED fits of this object thus suggest that it is not a typical high-redshift galaxy. Rather, its properties suggest either an ultra-faint, dust-obscured high-redshift quasar (or an object in transition between a normal high-redshift galaxy and an AGN-dominated one), or potentially an even more exotic compact object such as a clump of Pop.~III, supermassive or dark stars, or a direct-collapse black hole (DCBH), which we consider here as well. Note that we use here the term `quasar' in the simplest sense of an object with a quasi-stellar appearance, and in lenient alternation with `AGN', i.e. designating an object with a strong AGN contribution.

This paper is organized as follows: In \S \ref{sec:data} we describe the data used in this work. In \S \ref{sec:results} and \ref{sec:discussion} we report the discovery of the object and discuss its physical properties. The work is concluded in \S \ref{sec:conclusion}. Throughout this paper, we use a standard flat $\Lambda$CDM cosmology with $H_0=70\,\frac{\mathrm{km}}{\mathrm{s}\,\mathrm{Mpc}}$, $\Omega_{\Lambda}=0.7$, and $\Omega_\mathrm{m}=0.3$. Magnitudes quoted are in the AB system \citep{Oke1983ABandStandards} and all quoted uncertainties represent $1\sigma$ ranges unless stated otherwise. 

\begin{figure*}
    \centering
    \includegraphics[width=\textwidth]{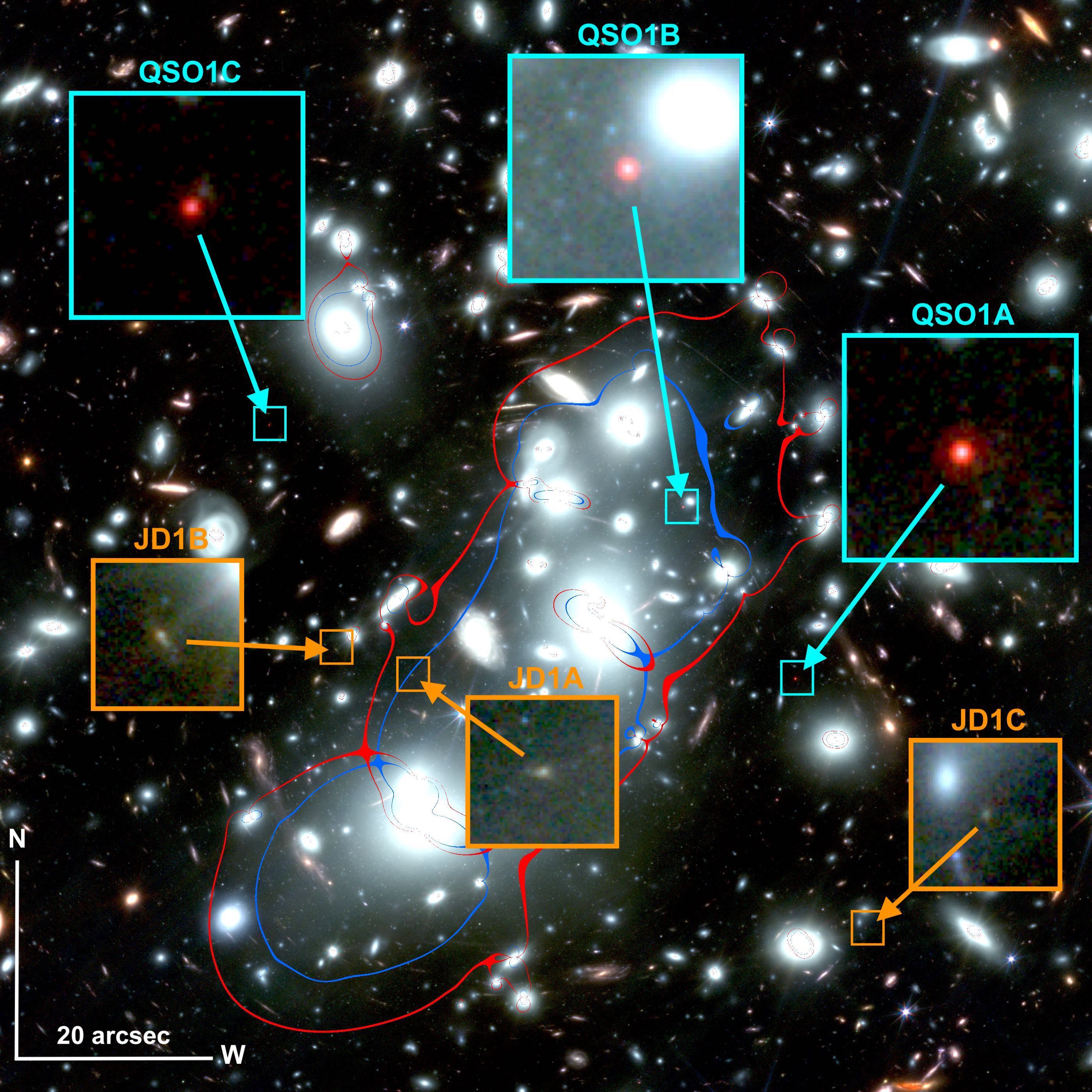}
    \caption{JWST/NIRCam composite-color image of Abell~2744 (Blue: F115W+F150W, Green: F200W+F277W, Red: F356W+F410W+F444W; see Fig.~4 in \citealt{Furtak2022UNCOVER} for a larger field-of-view) together with the critical curves from our SL model for sources at $z_{\mathrm{s}}=1.69$ (\emph{blue}) and $z_{\mathrm{s}}=7.5$ (\emph{red}). The notable three point-like red multiple images in \emph{cyan} squares are the images (and zoomed-in versions) of the quasar candidate reported here at a redshift of $z_{\mathrm{phot}}\simeq7.6$. We refer to these as \textit{QSO1A, B} and \textit{C}. For comparison, we also show the three images (JD1A, B and candidate C image) of the \citet{Zitrin2014highz} $z\simeq9.76$ object recently confirmed in \citet{Roberts-Borsani2022arXiv221015639R} with JWST spectroscopy, in \emph{orange} squares. All squares in the figure encompass about $2.4\arcsec\times2.4\arcsec$, and a 20\arcsec\ bar is given in the bottom left corner. Note that despite being at a lower redshift than JD, the faint quasar candidate is significantly redder. Also, the images of JD1 show significant structure, whereas the images of our object are point-like, implying a very compact de-magnified size of $r_{e}<35$\,pc, at most. The SED of the red object is shown in Figs.~\ref{fig:SED-galaxies} and~\ref{fig:SED-AGN}.}
    \label{fig:cc}
\end{figure*}

\section{Data}\label{sec:data}
The main data set used in this work are the recent observations taken with the \textit{Near Infrared Camera} \citep[NIRCam;][]{rieke05,Rieke2023} aboard JWST, carried out in the framework of the \textit{Ultra-deep NIRSpec and NIRCam ObserVations before the Epoch of Reionization} (UNCOVER) program \citep[Program ID: GO~02561; PIs: I.~Labb\'{e} \& R.~Bezanson;][]{Bezanson2022arXiv221204026B}. The JWST UNCOVER program was designed to observe the strong lensing (SL) galaxy cluster Abell~2744 (A2744 hereafter) with NIRCam to unprecedented depths in the F115W, F150W, F200W, F277W, F356W, F410M and F444W bands and over a large area of $\sim45$\,arcmin$^{2}$ around the cluster. A2744 has also recently been observed in two other JWST programs: The early release (ERS) survey GLASS-JWST \citep[Program ID: ERS~1324; PI: T.~Treu;][]{treu22}, and a director's discretionary time (DDT) program targeting a lensed supernova (Program ID: DD~2756; PI: P.~Kelly). The NIRCam imaging from these programs is also folded into the UNCOVER mosaics used in this work. The data were reduced and drizzled into mosaics with the \texttt{Grism redshift and line analysis software for space-based spectroscopy} \citep[\texttt{grizli}\footnote{\url{https://github.com/gbrammer/grizli}};][]{grizli2022}. The final mosaics used here have pixel scales of 0.02\arcsec\ and 0.04\arcsec\ in the short- and long-wavelength channels (SW and LW), respectively, and sum up $\sim16$\,h exposure time per filter on average, achieving $5\sigma$-depths of $29-30$\,AB magnitudes, or $\sim32$\,AB with lensing magnification. A photometric catalog of objects detected in the whole UNCOVER field is also generated in the UNCOVER program and is released with its other data products \citep{Weaver2023}. The catalog includes photometric redshift estimates for each identified object and lists the lensing magnifications from the UNCOVER lensing model \citep{Furtak2022UNCOVER} which is described in more detail in section~\ref{sec:SL}. We refer the reader to \citet{Bezanson2022arXiv221204026B} and \citet{Weaver2023} for more details of the data reduction, field depth in the different bands, source extraction and catalog assembly. All UNCOVER data products are publicly available on the UNCOVER website\footnote{\url{https://jwst-uncover.github.io/##releases}}. 

Note that the GLASS-JWST program also obtained \textit{Near-Infrared Imager and Slitless Spectrograph} \citep[NIRISS;][Doyon et al. in prep.]{doyon12} spectroscopy of the cluster core covering wavelengths $\lambda\sim1.1-2.2\,\mu$m. The UNCOVER program also includes spectroscopic follow-up observations with JWST's \textit{Near Infrared Spectrograph} \citep[NIRSpec;][]{jakobsen22,Ferruit2022,Boeker2023} which are scheduled for July~2023 and would allow deeper spectroscopic measurements and to longer wavelengths. Abell 2744 will also be observed with NIRCam as part of the JWST guaranteed time observation (GTO) program \textit{Prime Extra-galactic Areas for Reionization and Lensing Science} \citep[PEARLS; Program ID: GTO~1176; PI: R.~Windhorst;][]{windhorst2022}.

Ancillary data for A2744 include deep ($\sim29$\,AB) HST imaging, taken in various programs, most prominently the HFF, and HST grism spectroscopy from \textit{Grism Lens-Amplified Survey from Space} survey \citep[GLASS; PI: T.~Treu;][]{treu15}. We also note that spectroscopy with the \textit{Multi-Unit Spectroscopic Explorer} \citep[MUSE;][]{Bacon2010MUSE} on ESO's \textit{Very Large Telescope} (VLT) were used to measure the redshift for many multiply imaged galaxies in the cluster \citep{Mahler2018A2744,bergamini22}. A2744 is also part of the \textit{ALMA Frontier Fields} survey \citep[][]{gonzales-lopez17,Kohno2019asrc.confE..64K,Sun2022ApJ...932...77S} with the \textit{Atacama Large Millimeter/sub-millimeter Array} (ALMA), the band~6 ($\lambda\sim1.2$\,mm) imaging of which are also used in this work. Finally, A2744 has various X-ray observations \citep[see e.g.][]{Merten2011,Bogdan2023} which are publicly available on the \textit{Chandra} Data Archive and which we use to search for X-ray counterparts of the red object.

\section{A triply imaged compact object at $\lowercase{z}_{\mathrm{\lowercase{phot}}}\simeq7.6$} \label{sec:results}
We construct a composite-color image from the UNCOVER NIRCam mosaics including the previous programs mentioned in section~\ref{sec:data}. Three distinct red point-like objects stand out in the central core of the cluster (see Fig.~\ref{fig:cc}). These follow the lensing symmetry observed for other lensed systems and our lens model  \citep{Furtak2022UNCOVER} indeed predicts these to be counter images of the same object at a high redshift, namely $z_{\mathrm{model}}\gtrsim7$ (but see \ref{sec:SL} for more details). Two of the images reported here were in fact already designated as high-redshift $z\sim7-8$ multiple-image candidates by \cite{Atek2014A2744,atek15a} in the HFF data \citep[see also][]{Lam2014modelA2744} and were included in the \citet{Mahler2018A2744} SL model of A2744. The recent JWST data now not only allow us to detect the third counter image but also reveal the peculiar properties of this source. In the following we review the observed and physical properties of the red object and discuss its possible origins.

\subsection{Photometry} \label{sec:photometry}

\begin{splitdeluxetable*}{lccccccccBcccccccccc}
\label{tab:photometry}
\tablecaption{Photometry of the three multiple images of the red compact object.}
\tablewidth{\textwidth}
\tablehead{
\colhead{ID} & \colhead{R.A.} & \colhead{Dec.}  & \colhead{F435W} & \colhead{F606W} &\colhead{F814W}  &\colhead{F105W} & \colhead{F115W} &\colhead{F125W} & \colhead{F140W} &\colhead{F150W} & \colhead{F160W} &\colhead{F200W} & \colhead{F277W} &\colhead{F356W} & \colhead{F410W} &\colhead{F444W} & \colhead{ALMA Band~6} & \colhead{$\mu$}
}
\startdata
Image~A~(8296)    &  00:14:19.161   &   -30:24:05.664 &   $<4$ &  $<2$     &  $<2$     &  $16\pm2$    &   $22\pm2$    &   $23\pm3$    &   $26\pm3$    &   $24\pm2$    &   $32\pm3$    &   $28\pm2$    &   $32\pm1$    &   $167\pm1$   &   $283\pm3$   &   $359\pm2$   &   $<85.5\times10^{3}$ &   $7.5\pm0.4$\\
Image~B~(9992)    &   00:14:20.051  &   -30:23:48.058  &   $<4$ &  $<2$ &  $<2$ &  $18\pm2$   &   $20\pm2$   &   $26\pm2$   &   $22\pm3$   &   $27\pm3$   &   $22\pm3$   &   $29\pm3$   &   $35\pm2$   &   $130\pm1$   &   $212\pm1$   &   $262\pm1$   &   $<85.5\times10^{3}$&   $8.4\pm0.8$\\
Image~C~(10712)   &   00:14:23.331   &   -30:23:39.639  &   $<4$ &  $<2$     &  $<2$  &  $5\pm1$    &   $8\pm1$    &   $9\pm1$    &   $7\pm1$    &   $10\pm1$    &   $6\pm1$    &   $9\pm1$    &   $12\pm1$    &   $49\pm1$    &   $79\pm1$    &   $98\pm1$   &   $<67.7\times10^{3}$&   $4.0\pm0.1$\\
\enddata
\tablecomments{Fluxes are in nJy from the deep HST/\textit{Advanced Camera for Surveys} (ACS), HST/\textit{Wide-Field Camera Three} (WFC3) and JWST/NIRCam imaging. For image A fluxes are taken from the UNCOVER photometric catalog \citep[][0.32\arcsec\ apertures]{Weaver2023}. For images B and C, which are near a cluster galaxy and a foregound blue galaxy, respectively, we use a dedicated PSF-corrected aperture photometry measurements. Note that the fluxes listed here are \textit{not} yet corrected for magnification. For each of the three images we also indicate the catalog ID number in parentheses. For bands in which the images are not detected, we list the $1\sigma$-upper limit. The magnifications are computed from our UNCOVER-based SL model of A2744 published in \citet{Furtak2022UNCOVER}.}
\end{splitdeluxetable*}

The three images of our object are detected in the UNCOVER catalog \citep[][for details]{Weaver2023}. The detection is performed on the UNCOVER mosaics corrected for intra-cluster light (ICL) with the python implementation of \texttt{SExtractor} \citep[\texttt{SEP};][]{BertinArnouts1996Sextractor,barbary16,barabry18}. We note however that for images B and C, the UNCOVER catalog fluxes are heavily contaminated with light from the nearby cluster galaxy in image B (see Fig.~\ref{fig:cc}) and from a faint blue object right next to image C (also visible in Fig.~\ref{fig:cc}). In order to derive de-contaminated fluxes for these two images, we measure them in a circular aperture of 0.32\arcsec\ diameter using \texttt{photutils} \citep[\texttt{v1.6.0};][]{photutils22} and subtract a local background measured in an annulus around the aperture. The background annuli are carefully optimized to not over-subtract the background in image B and to exclude the blue neighboring object in image C. The fluxes are then aperture corrected using the point-spread-function (PSF) in each band. For consistency, we compare the fluxes of image A measured with our method to the UNCOVER catalogs results and find that they concur well. The photometry of all three images used in this work is listed in Tab.~\ref{tab:photometry}. The flux ratios of images B and C to image A are slightly systematically lower than -- but broadly consistent with -- the magnification ratios predicted by the lens model (see section~\ref{sec:SL}). Some color discrepancies are seen, in particular in image B which we note may still contain residual ICL contamination. However, we find that these discrepancies are below the expected systematics from the background estimation, probed by examining different background annulii. We can only conclude therefore that the colors and flux levels of the images are broadly consistent, also with the lens model prediction, but do not at this stage detect a significant color variation between the three images.

The UNCOVER catalog also contains a first photometric redshift estimate of $z_{\mathrm{phot}}=7.188_{-0.042}^{+0.006}$ computed with \texttt{EAZY} \citep{Brammer2008EAZY}. This agrees with the high-redshift geometric estimate from our lens model (see section~\ref{sec:SL}). We note however that \texttt{EAZY} does not yield a good fit for this object (e.g. $\chi^2=462$ for image A). This is due to the extremely red colors that this object has in the JWST filters (see section~\ref{sec:color}) which cannot be properly reproduced with the standard template set used with \texttt{EAZY} for the UNCOVER catalogs (unless a combination of several galaxy templates are used), and hints at its peculiar nature. Note that AGN templates are not included in the \texttt{EAZY} run.

While this object is covered by the GLASS-JWST NIRISS observations (see section~\ref{sec:data}), none of the three images is detected spectroscopically. This is not surprising however since the three images are relatively faint ($\sim28$\,AB magnitudes) in the $\lambda\sim1.1-2.2\,\mu$m range covered with NIRISS. The strong emission in the redder bands F356W, F410M and F444W (down to $\sim25$\,AB) will however be picked up in the planned UNCOVER JWST/NIRSpec observations which will reach 5$\sigma$ continuum depths of $28-29$\,AB magnitudes. These correspond to limiting $3-4\sigma$ emission line sensitivities of $0.5-1\times10^{-19}\,\frac{\mathrm{erg}}{\mathrm{s\,cm}^2}$ in the F444W-band assuming an intrinsic line width of $500\,\frac{\mathrm{km}}{\mathrm{s}}$ at $z=7.6$ computed with the JWST/NIRSpec exposure time calculator (ETC).

We also note that the object is not detected in the ALMA data (see Tab.~\ref{tab:photometry} for upper limits and section~\ref{sec:data} for details), nor do we observe any counterparts in the 1.25\,Ms of recently acquired \emph{Chandra} X-ray data \citep{Bogdan2023}.

\subsection{Colors} \label{sec:color}

\begin{figure}
    \centering
    \includegraphics[width=\columnwidth]{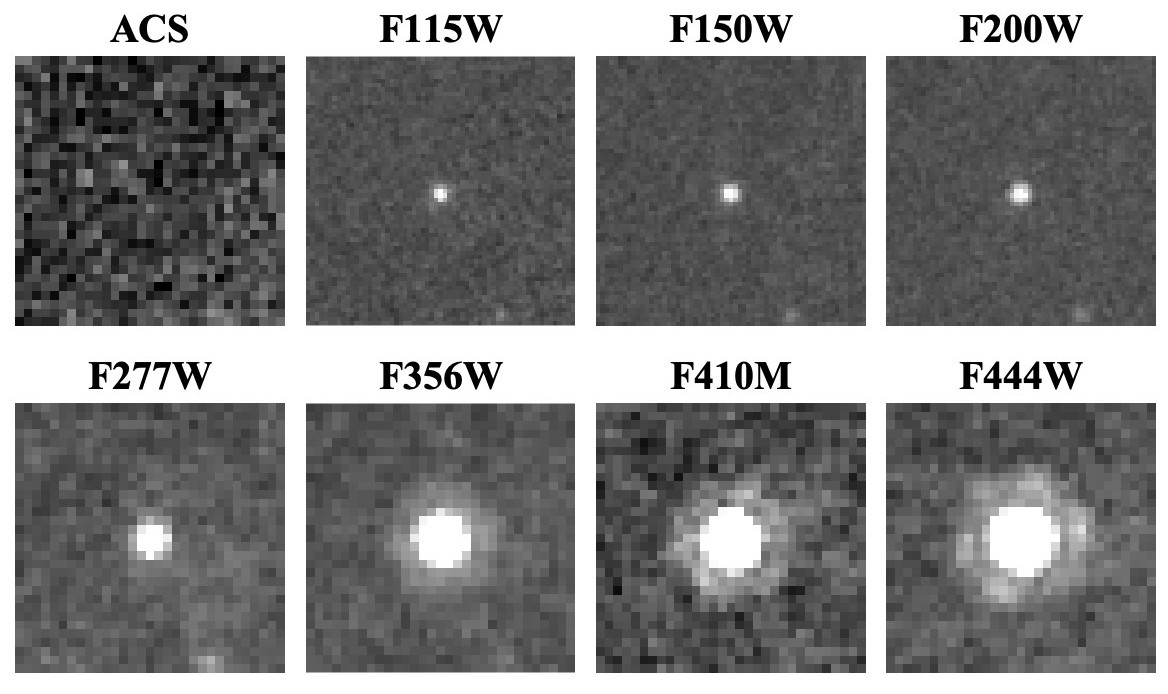}
    \includegraphics[width=\columnwidth]{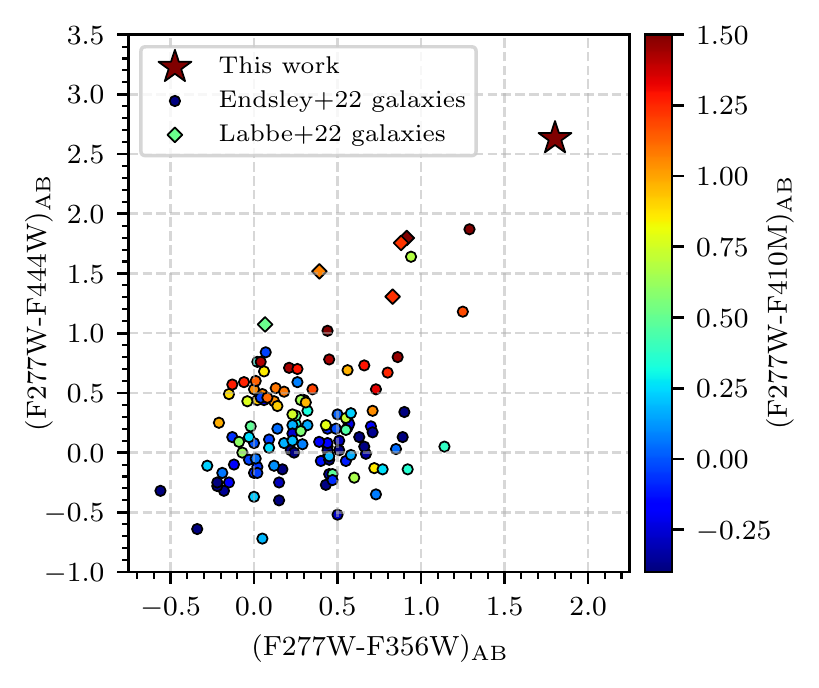}
    \caption{\textit{Upper panel}: Our red compact object in the different JWST/NIRCam bands taken with UNCOVER. The upper left panel shows a stack of the HFF ACS which shows a clear non-detection. Each square is $1.2\arcsec\times1.2\arcsec$. We show here the data for the first multiple image (image~A). \textit{Bottom panel}: Color-color diagram of our object (star) with typical NIRCam-detected $z\sim7-8$ galaxies \citep{Endsley2022arXiv220814999E} (circles) and red $z\sim9$ galaxies \citep[][]{Labbe2022arXiv220712446L,Finkelstein2022arXiv221105792F} (squares) in the CEERS field. The source presented in this work is redder by about a magnitude compared to the reddest of the CEERS sources, which hints at heavy emission line activity or a very strong and dusty rest-frame optical continuum.}
    \label{fig:colorcolor}
\end{figure}

The object appears to be very red, with NIRCam colors, in magnitudes:

\begin{align*}
    \mathrm{F115W}-\mathrm{F150W}&=0.12\pm0.13\\
    \mathrm{F150W}-\mathrm{F200W}&=0.14\pm0.11\\
    \mathrm{F200W}-\mathrm{F277W}&=0.15\pm0.08\\
    \mathrm{F277W}-\mathrm{F356W}&=1.80\pm0.10\\
    \mathrm{F277W}-\mathrm{F410M}&=2.37\pm0.05\\
    \mathrm{F277W}-\mathrm{F444W}&=2.63\pm0.10\\
    \mathrm{F356W}-\mathrm{F410M}&=0.57\pm0.10\\
    \mathrm{F356W}-\mathrm{F444W}&=0.83\pm0.08\\
\end{align*}

At $z\sim7.5-8$ the H$\beta$ and [\ion{O}{3}] lines are shifted into the F410M- and F444W-bands, whereas the [\ion{O}{2}] doublet is shifted into the F356W-band, which may account for the very red colors. Indeed, these lines have been observed to be sufficiently strong to modify and boost the broadband colors of galaxies. In fact, they have also been used as indicators for possible Lyman-$\alpha$ emitters (LAEs) in the EoR \citep[e.g.][]{Labbe2013,Smit2015ApJ...801..122S,Roberts_Borsani2015,Stark2017}. However, the NIRCam colors that we find here seem to be much redder than expected for typical early galaxies at $z\sim7-8$. In Fig.~\ref{fig:colorcolor} we show color-color diagrams demonstrating that our red object is indeed extreme: it clearly falls outside the region in which typical galaxies found at similar redshifts with JWST reside \citep{Endsley2022arXiv220814999E} and in the direction which suggests higher emission lines, dust extinction, and/or a strong Balmer break. For comparison, \citet{Endsley2022arXiv220814999E} have found that one red galaxy in their $z\sim7-8$ sample (Fig.~14 in \citealt{Endsley2022arXiv220814999E}, see also \citealt{Labbe2022arXiv220712446L}) can also be explained explained by a combination of a galaxy with an AGN, resulting in an [\ion{O}{3}]+H$\beta$ equivalent width (EW) of over 5000\,\AA. Our object has even more extreme colors and can thus be expected to either have even stronger emission lines, which might suggest an AGN component, or an extremely red and dusty rest-frame optical continuum. Both scenarios will be further discussed in the following sections. Note that our object is also significantly redder than the red galaxy population detected by \citet{Labbe2022arXiv220712446L} at higher redshifts (also shown in Fig.~\ref{fig:colorcolor}).

\subsection{Spectral slopes} \label{sec:slope}
We fit the continuum of our object with a power-law $F_{\lambda}\propto\lambda^{\beta}$. The UV-continuum is estimated in the rest-frame 1300\,\AA\ to 2300\,\AA\ range (i.e. the F115W through F200W bands) and we obtain a UV-slope of $\beta_{\mathrm{UV}}=-1.6\pm0.2$. 

For comparison, the UV-slope of our object is somewhat bluer than that of a red quasar, such as e.g. the $z\sim7$ object found by \citet{Fujimoto2022Natur.604..261F}, but typical of un-obscured AGN \citep[e.g.][]{VandenBerk2001,Selsing2016}. This slope is also of the same order as other high-redshift galaxies with similar absolute UV luminosities \citep{Bouwens2014UVslopes}. It is however notably redder than typical early galaxies recently detected in JWST observations \citep[e.g.][see also discussion in \citealt{Endsley2022arXiv220814999E}]{Atek2023,Adams2023MNRAS.518.4755A,Castellano2022arXiv221206666C,Cullen2022UVslopes,furtak22b}, or the population of galaxies at high redshifts with blue UV-slopes but red rest-frame optical colors mentioned above \citep{Labbe2022arXiv220712446L,Finkelstein2022arXiv221105792F}.

We also estimate the optical continuum slope, in the rest-frame 3200\,\AA\ to 5100\,\AA\ range (F277W through F444W). We find a much redder optical slope of $\beta_{\mathrm{optical}}=1.89\pm0.04$ compared to the UV slope, in line with the extremely red colors seen for this object towards longer wavelengths.

%Some of the objects in this sample show red colors especially in the LW channels (e.g. $\mathrm{F277W}–\mathrm{F444W}\sim2$), somewhat similar to our object (see Fig.~\ref{fig:colorcolor}). However, as we show in sections~\ref{sec:size} and~\ref{sec:UV-density}, our object is distinct with respect to both typical blue high-redshift galaxies, and this population of red galaxies, in terms of size compared to its UV luminosity (see section~\ref{sec:luminosity}).}

\subsection{Lensing magnifications and geometric redshift} \label{sec:SL}
To estimate the de-lensed properties of the object presented here as well as to corroborate the photometric redshift estimates, we use an updated fully parametric SL model of A2744 that we recently constructed in the framework of the UNCOVER program \citep[][see also \citealt{Roberts-Borsani2022arXiv221015639R}]{Furtak2022UNCOVER}. The model uses a long list of spectroscopically confirmed multiply imaged systems \citep[e.g.][]{Mahler2018A2744,bergamini22} and additional photometric systems identified in the new UNCOVER data, in particular in the two northern sub-clusters. Note that this sub-structure only has a minor effect on the SL properties in the main cluster core though, where the object studied in this work is situated. For the lens model, cluster galaxies are modeled each as a dual pseudo-isothermal elliptical mass distribution (dPIE; see \citealt{Keeton2001models,Eliasdottir2007arXiv0710.5636E}) and five dark matter (DM) halos are used, each modeled as a pseudo-isothermal elliptical mass distribution  \citep[PIEMD; e.g.][]{Keeton2001models}. The final model used in this work reproduces the $>60$ multiple image systems which span a wide redshift range from $z\sim1.7$ to $z\simeq10$ very well with a lens plane RMS of $0.66\arcsec$. Full details of the model are given in \citet{Furtak2022UNCOVER}. The critical curves from the model for various redshifts can be seen in Fig.~\ref{fig:cc}.

We use this SL model to obtain magnifications of $\mu=7.5\pm0.4$, $\mu=8.4\pm0.8$ and $\mu=4.0\pm0.1$ for the images A, B, and C, respectively, assuming a source redshift of $z_{\mathrm{s}}\sim7.5$. These are in broad agreement with the flux ratios that we measure for the three objects (see Tab.~\ref{tab:photometry}). 

We can also use the SL model to examine the redshift of the source, due to the nesting effect in which the lensing critical curves grow for higher source redshifts. However, as the angular diameter distance ratio saturates for high-redshift sources, a concrete redshift estimate is not easy to obtain, but a lower limit can be placed. In particular, our model from \citet{Furtak2022UNCOVER} strongly suggests that the object lies at $z>7$ and in fact pushes to even higher redshifts around $z\sim10-11$ (although these higher redshifts are ruled out by the photometry). To further examine this we therefore generate a suite of different models spanning a larger range of input configurations (i.e. number of DM halos, number of freely weighted massive cluster galaxies, number of photometric multiply imaged systems and multiple image configurations). These models explicitly include the red object system with a free redshift to be optimized and also two new dropout systems at $z_{\mathrm{phot}}\simeq4.9$ and $z_{\mathrm{phot}}\simeq6.9$ identified in the UNCOVER data close to the images of the red object. The various resulting models span a range of best-fit redshift estimates for the red object system, with the lowest-redshift ones being $z\sim5.5$. A similar constraint can in fact be also obtained model-independently from the geometry of multiple image systems near the red object: Systems up to $z\sim5$ are seen as expected at somewhat smaller angles than the red object, implying that the red object system should lie at a higher redshift.

\subsection{Luminosity} \label{sec:luminosity}
Extrapolating the fitted UV-continuum to 1450\,\AA, we measure a relatively faint UV luminosity of $M_{\mathrm{UV},1450}=-16.81\pm0.09$ after correcting for gravitational magnification.

We furthermore compute a lower limit estimate of the bolometric luminosity by integrating the broad-band photometry in all available bands (see Tab.~\ref{tab:photometry}) over their respective bandwidths which yields $L_{\mathrm{bol}}>2\times10^{43}\,\frac{\mathrm{erg}}{\mathrm{s}}$ after magnification correction. Note that this only constitutes a lower limit on the bolometric luminosity because our measurements only span the rest-frame UV and optical spectral ranges. Another estimate for the bolometric luminosity is given in section \ref{sec:sed} based on the SED fit.

Finally, the ALMA non-detection of our object results in an upper limit on the rest-frame infrared (IR) luminosity of $L_{\mathrm{IR}}<2\times10^{10}\,\mathrm{L}_{\odot}$ at $1\sigma$, assuming a single modified black-body with a dust temperature $T_{\mathrm{d}}=47\,\mathrm{K}$ and a dust-emissivity spectral index $\beta_{\mathrm{d}}=1.6$ \citep{Beelen2006} scaled to the ALMA non-detection as in e.g. \citet{Furtak2023AGN}.

\subsection{Size}\label{sec:size}
We verify that the source is indeed an unresolved point source with a dedicated \texttt{GALFIT} \citep{Peng2010AJ....139.2097P} analysis in all available filters. From the \texttt{GALFIT} measurement, we obtain an observed effective radius of $r_e\simeq1.3$\,pixels (with a pixel scale of 0.02\arcsec/pixel) in the SW bands for both of the images that are not heavily affected by a nearby cluster galaxy light (i.e images A and C, see Fig.~\ref{fig:cc} and section~\ref{sec:photometry}), and $r_e<1$\,pixel in the LW bands. The typical statistical error for these fits in both the SW and LW channels is $\lesssim0.05$ respective pixels. Taking the magnification into account, this size estimate translates into a source radius upper limit of $r_e\lesssim100$\,pc. Moreover, we note that both image A and image C have a very similar effective radius fit of $r_e\simeq1.3$\,pixels. However, by the magnification ratio, we would expect image A to have a 40\% larger radius if the images were even marginally resolved. This suggests that indeed, the object is an unresolved point source.

Given the point-like nature of the source, assuming a PSF with a full-width half maximum (FWHM) of 0.04\arcsec\ for the shortest-wavelength filter and taking the magnification into account, we place an upper-limit of FHWM$\lesssim70$\,pc on the source. The relation between the FWHM and effective radius is not straight forward and depends on the light profile shape of the source \citep{VoigtBridle2010MNRAS.404..458V,Ryon2017ApJ...841...92R}. For simplicity, we here assume the effective radius $r_{e}$ to be half of the FWHM. Note that this is somewhat conservative, e.g. for an exponential disk the factor is higher, $r_{e}\sim\frac{\mathrm{FWHM}}{2.43}$ \citep{Murphy2017ApJ...839...35M}. We thus obtain an upper limit on the source size of $r_{e}\lesssim35$\,pc.

This limits the possible nature of the source somewhat, by largely ruling out a massive and extended evolved galaxy in which the red colors would originate from stellar continuum alone. Instead, if the emission of our object is indeed dominated by starlight, it would need to be a bright and dense clump of stars (see also discussion in section~\ref{sec:discussion}). These are however usually close to a host galaxy which is clearly not observed here. Indeed, the sizes of lensed high-redshift galaxies measured with HST imply very small $r_e\sim$100\,pc scale upper limits \citep[e.g.][]{Coe2012highz,Zitrin2014highz,Bouwens2017,Kawamata2018ApJ...855....4K}.The scenario in which only a bright clump from a distant galaxy is seen, i.e. the host galaxy is below the detection limit, therefore needs to be considered. Observations of high-redshift galaxies with JWST have already 
revealed them to show sub-structures consistent with multiple star-forming clumps, as e.g. in the multiply imaged $z\sim10$ object in A2744 \citep{Zitrin2014highz,Roberts-Borsani2022arXiv221015639R} shown in Fig.~\ref{fig:cc} or in the $z\sim10$ object recently observed by \citet{Hsiao2022arXiv221014123H}. With the JWST UNCOVER data, which are much deeper than other JWST images so far, and despite the relatively high lensing magnification (see Tab.~\ref{tab:photometry}), we do not find any hint of an underlying more extended structure to our object. This implies that the measured luminosity may indeed arrive from a single central source with a size on the scale of at most several tens of pc.

\subsection{Luminosity density} \label{sec:UV-density}

\begin{figure}
    \centering
    \includegraphics[width=\columnwidth]{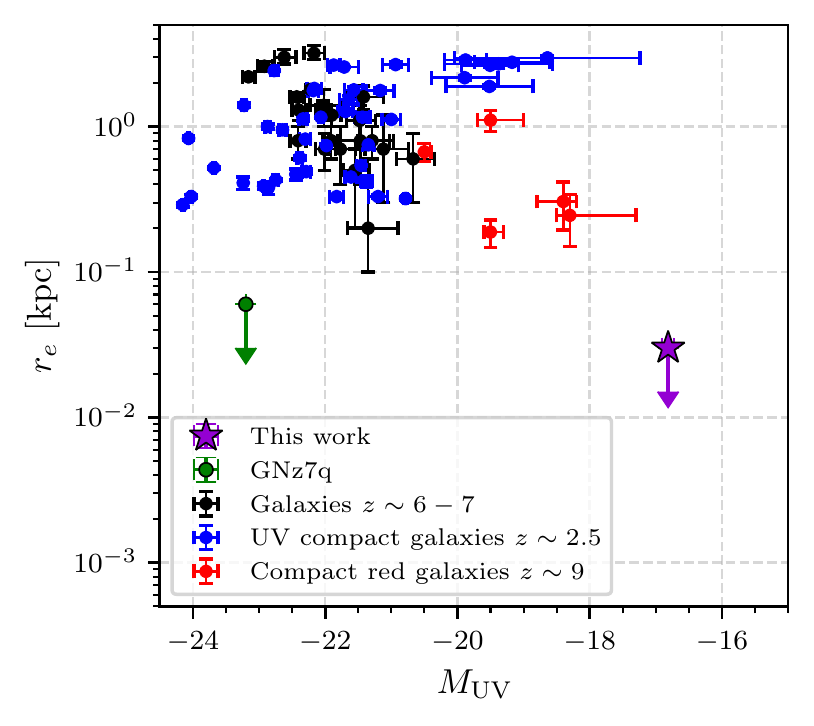}
    \caption{$M_{\mathrm{UV}}$-size diagram. Our red compact object is shown as the purple star. For comparison we also show typical compact UV-emitting galaxies at both low redshifts (from \citealt{barro2014}; blue dots), and high redshifts (from \citealt{bowler2017}; black dots). We also show red galaxies recently detected with JWST in the CEERS field by both \citet{Labbe2022arXiv220712446L} and \citet{Finkelstein2022arXiv221105792F}. Our red object is more compact by about an order of magnitude compared to the smallest of these galaxies. Also shown in green is an obscured quasar, or a hybrid galaxy-AGN candidate, GNz7q, recently detected by \citet{Fujimoto2022Natur.604..261F}.}
    \label{fig:MUV-size}
\end{figure}

In Fig.~\ref{fig:MUV-size}, we put the UV luminosity measured in section~\ref{sec:luminosity} in relation to our source-size upper limit measured in section~\ref{sec:size}. The red compact object studied here seems to have a very high UV luminosity density, occupying a different region of the size-luminosity diagram than most typical galaxies both at high and low redshifts. In particular it is smaller by about an order of magnitude than the smallest of these galaxies. This implies that it may indeed be too bright for its size to be explained by regular stars alone. For reference, our object has a UV luminosity density of at least an order of magnitude higher than most of the \citet{Labbe2022arXiv220712446L} red high-redshift galaxies and of similar order as the very brightest ($M_{\mathrm{UV}}\sim-24$) and most compact ($r_e\sim300$\,pc) low-redshift UV emitters. Bearing in mind that the object is not a transient source, e.g. it was already detected in the HFF (see also our discussion of time delays in section~\ref{sec:discussion}), only few possible scenarios remain. For example, the main engine driving the observed luminosity and colors could possibly be a faint and dust-obscured AGN such as in e.g. \citet{Fujimoto2022Natur.604..261F} or one of the objects in \citet{Endsley2022arXiv220814999E}, or possibly a more exotic source such as a compact clump of Pop.~III or supermassive stars (although the object does not seem to show colors blue enough to support the former, as discussed in section~\ref{sec:discussion}). While the current data set in absence of spectroscopy does not allow a robust discussion of the possible scenarios, we nevertheless perform an SED analysis in section~\ref{sec:sed}, showing that the AGN scenario is plausible. A more detailed source analysis is deferred to future work when the planned UNCOVER JWST/NIRSpec observations are available \citep[scheduled for July~2023;][]{Bezanson2022arXiv221204026B}.

\subsection{SED fitting}\label{sec:sed}
In order to further investigate the origin of the object's observed properties and to constrain its parameters, we perform several SED fits to the de-magnified photometry of image~A (see Tab.~\ref{tab:photometry}), using the \texttt{BayEsian Analysis of GaLaxy sEds} tool \citep[\texttt{BEAGLE};][]{chevallard16}, \texttt{Prospector} \citep{prospector21} and the \texttt{Code Investigating GALaxy Emission} \citep[\texttt{CIGALE};][]{Boquien2019,Yang2020CIGALE1,Yang2022CIGALE2}. We first fit our object with a star-forming galaxy in section~\ref{sec:galaxy-SEDs} and then add an AGN component in section~\ref{sec:AGN-SEDs}. Finally, we fit a custom model of obscured AGN SED in section~\ref{sec:Ivo-SEDs}.

\subsubsection{Galaxy SED fits} \label{sec:galaxy-SEDs}

\begin{figure*}
    \centering
    \includegraphics[width=\textwidth]{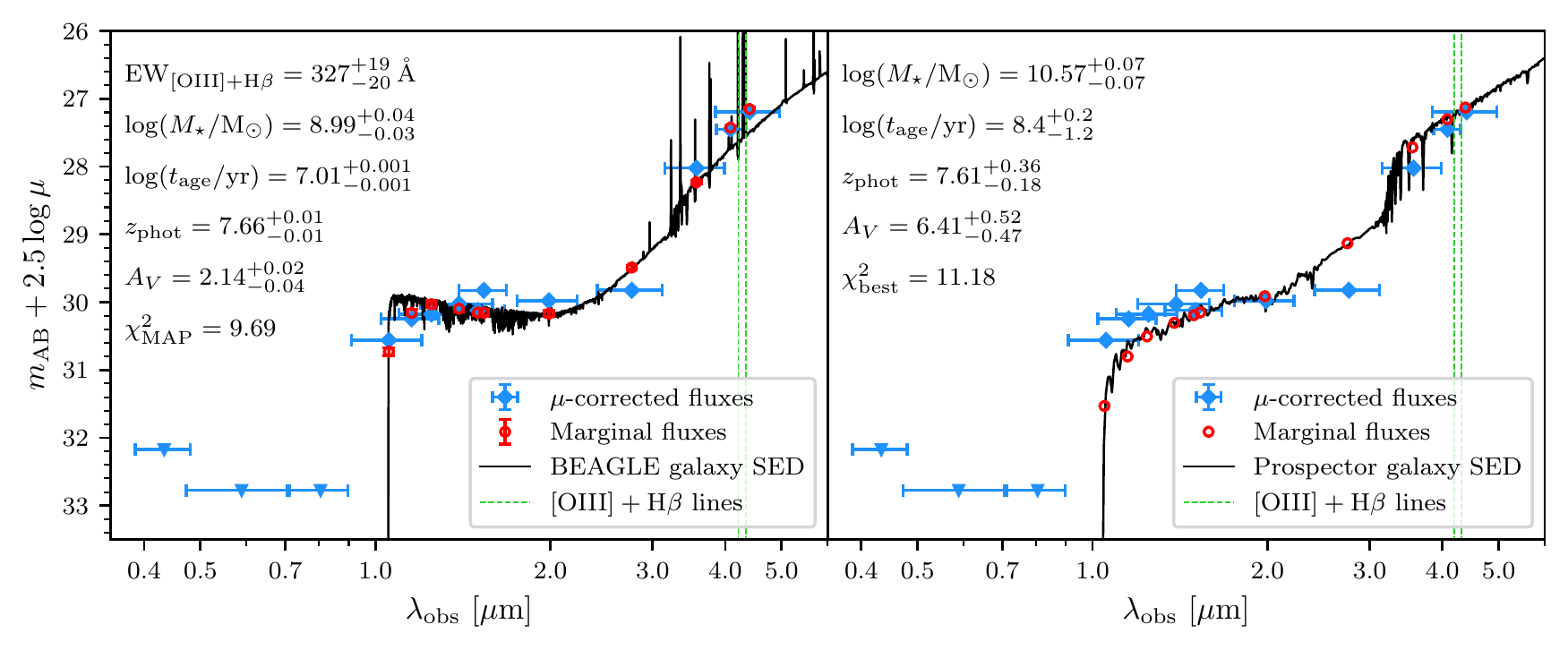}
    \caption{Maximum-a-posteriori (MAP, i.e. best-fit) galaxy SEDs fitted with \texttt{BEAGLE} and \texttt{Prospector} (black). The de-magnified photometry of image~A is shown in blue, and the marginal model fluxes in each band calculated from the posterior distribution are shown in red. \textit{Left}: \texttt{BEAGLE} SED fit with a constant SFH star-forming galaxy. \textit{Right}: \texttt{Prospector} SED fit with a non-parametric SFH star-forming galaxy. Both fits consistently find our object to lie at a redshift of $z_{\mathrm{phot}}\sim7.6-7.7$. The expected position of the rest-frame optical [\ion{O}{3}]$\lambda5007$\AA\ and H$\beta$ lines at the best-fit photometric redshift are indicated in green. Also noted on the figures are the resulting physical parameters and reduced $\chi^{2}$ values.}
    \label{fig:SED-galaxies}
\end{figure*}

For the galaxy SED fit with \texttt{BEAGLE}, we follow \citet{Endsley2022arXiv220814999E}, i.e. we adopt log-uniform priors on stellar mass ($\log(M_{\star}/\mathrm{M}_{\odot})\in[5, 12]$), age ($\log(t_{\mathrm{age}}/\mathrm{yr})\in[6, t_{\mathrm{Universe}}]$), metallicity ($\log(Z/\mathrm{Z}_{\odot})\in[-2.2, -0.3]$) and galaxy-wide effective ionisation parameter ($\log(\hat{U})\in[-4, -1]$). The dust-to-metal mass ratio is fixed to $\xi_{\mathrm{d}}=0.3$ and we adopt a uniform prior on the effective \textit{V}-band dust attenuation optical depth of $\hat{\tau}_V\in[0.001, 5]$. We use the standard \citet{gutkin16} set of \texttt{BEAGLE} templates which combine the latest version of the \citet{BC03} stellar population models with \texttt{Cloudy} \citep{ferland13} photoionization models to account for nebular emission. We assume a Chabrier initial stellar mass function (IMF) \citep{chabrier03}, a constant star-forming history (SFH) and an SMC dust attenuation law \citep{pei92}, and account for intergalactic medium (IGM) attenuation using the \citet{inoue14} IGM models.

This fit is complemented with an independent galaxy SED fit with \texttt{Prospector} \citep{prospector21}. We in particular use \texttt{Prospector}-$\alpha$ which models a non-parametric SFH and assumes a continuity prior to ensure smooth transitions between time bins \citep{Leja2017,Leja2019}. We additionally include two priors on the stellar mass and the SFH from \citet{Wang2023}. The stellar mass prior is constructed from the observed mass functions in \citealt{Leja2020} and the dynamic SFH prior is a simple phenomenological description reflecting the consistent observational finding that massive galaxies form much earlier than low-mass galaxies \citep{Cowie1996,Thomas2005}. The \texttt{Prospector} fit assumes a \citet{CF2000} dust attenuation in this work.

Both best-fitting galaxy SEDs are presented in Fig.~\ref{fig:SED-galaxies}. They both consistently find a very well constrained photometric redshift of $z_{\mathrm{phot}}\sim7.6-7.7$, i.e. somewhat higher than the \texttt{EAZY}-derived redshift from the UNCOVER catalog (see section~\ref{sec:photometry}). Both codes also find extremely large stellar masses of order $\log(M_{\star}/\mathrm{M}_{\odot})\sim9-11$ because they produce the measured red optical colors of our object (see section~\ref{sec:color}) with mostly stellar continuum. Such a massive galaxy is however very unlikely given the extremely compact nature of the source of $r_e\lesssim35$\,pc (see section~\ref{sec:size}): the resulting stellar mass density would exceed the densest known globular clusters by orders of magnitude. A more likely scenario, for example, could therefore be that the red optical colors are also driven by powerful nebular emission lines rather than stellar continuum alone, as investigated in section~\ref{sec:AGN-SEDs}. In addition, both galaxy SED fits seem to require large dust attenuations to produce the red optical continuum. The reader should bear in mind though that we do not detect any dust continuum emission in the ALMA 1.2\,mm imaging at $<85\,\mu$Jy (see Tab.~\ref{tab:photometry}) and that the $A_V=6.4\pm0.5$ found by \texttt{Prospector} in particular is not consistent with the upper limit on the IR luminosity estimated in section~\ref{sec:luminosity}, and the $A_V\sim2$ found by the \texttt{BEAGLE} fit, is only marginally consistent.

\subsubsection{Galaxy+AGN SED fits} \label{sec:AGN-SEDs}

\begin{figure*}
    \centering
    \includegraphics[width=\textwidth]{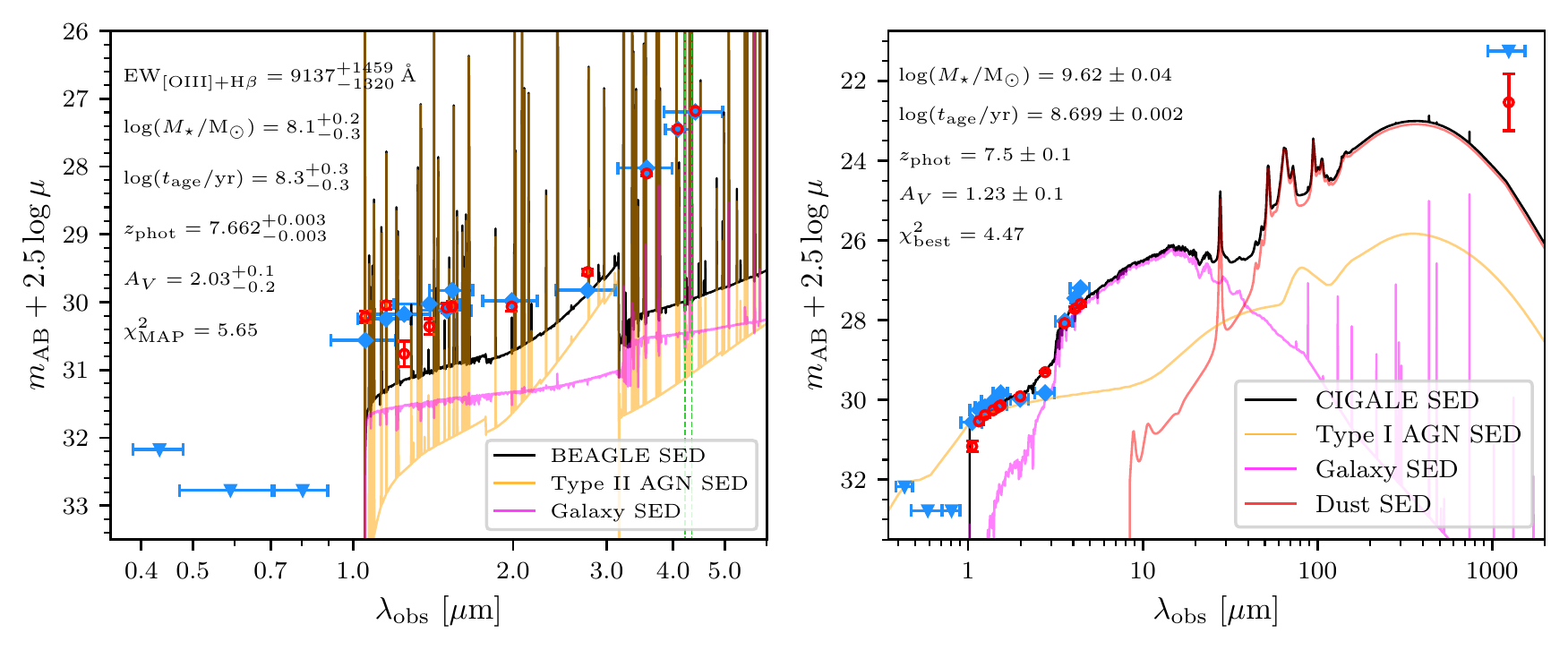}
    \caption{Maximum-a-posteriori (MAP) AGN SEDs fitted with \texttt{BEAGLE-AGN} and \texttt{CIGALE} (black). The de-magnified photometry of image~A is shown in blue, and the marginal fluxes in each band, computed from the posterior distribution, are shown in red as in Fig.~\ref{fig:SED-galaxies}. \textit{Left}: \texttt{BEAGLE-AGN} SED fit with a star-forming galaxy and a type~II AGN, similar to \citet{Endsley2022arXiv220814999E}. The expected position of the rest-frame optical [\ion{O}{3}]$\lambda5007$\AA\ and H$\beta$ lines at the best-fit photometric redshift are indicated in green. \textit{Right}: \texttt{CIGALE} SED fit with the combination of a galaxy with a type~I AGN and a dust emission component. We decompose the best-fitting SEDs (black) into their individual components, the galaxy SED (purple) and the AGN emission (orange) and the dust emission (red). The galaxy+AGN fits seems to better reproduce the observed red colors and red UV-slope (see section~\ref{sec:color}) than the galaxy-only fits shown in Fig.~\ref{fig:SED-galaxies}. This is also consistent with the luminosity density of our source described in sections~\ref{sec:size} and~\ref{sec:UV-density}. Also noted on the figures are the resulting physical parameters and reduced $\chi^{2}$ values. In the right-hand panel, the best-fit \texttt{CIGALE} SED (black) includes IGM attenuation while the individual components do not.}
    \label{fig:SED-AGN}
\end{figure*}

We use two codes that can also combine AGN emission models with that of the galaxy component: \texttt{BEAGLE-AGN} \citep[][]{vidal-garcia22}, which includes type~II AGN templates \citep[][]{feltre2016} incorporated into \texttt{BEAGLE} by E. Curtis-Lake to add AGN narrow-line emission to the galaxy templates; and \texttt{CIGALE}, which accommodates also a type~I AGN, which we \textit{a priori} expect to be the most consistent with the object's compact morphology (see section~\ref{sec:size}). For the \texttt{BEAGLE-AGN} fit, we adopt a similar configuration and parameter space as in section~\ref{sec:galaxy-SEDs}, again following \citet{Endsley2022arXiv220814999E}. For the \texttt{CIGALE} fit, we follow the configuration of \citet{Yang2023CEERSAGN}, assuming a delayed SFH, the \citet{BC03} stellar population models with a Chabrier IMF and stellar metallicity $Z=0.02\,\mathrm{Z}_{\odot}$, nebular emission models by \citet{Villa-Velez2021}, a modified \citet{Calzetti2000} dust attenuation law \citep{Leitherer2002}, the \citet{Draine2014} dust emission models for the host galaxy, the \texttt{Skirtor2016} \citep{Stalevski2012,Stalevski2016} 3D radiative transfer AGN models , which account for the AGN emission from the far UV to the torus and polar dust emission in the IR, and the \citet{Meiksin2006} IGM attenuation models. We refer the reader to \citet{Boquien2019} and \citet{Yang2020CIGALE1} for more details of the \texttt{CIGALE} templates used here. The AGN component in \texttt{CIGALE} can be of type~I or type~II, as determined through the inclination angle which we keep as a free parameter to allow for both options. Since \texttt{CIGALE} includes dust emission models, we also include the ALMA non-detection (see Tab.~\ref{tab:photometry}) as an upper limit in this fit.

The best-fitting SEDs of both galaxy+AGN fits are shown in Fig.~\ref{fig:SED-AGN}. Both fits yield photometric redshifts of $z_{\mathrm{phot}}\sim7.6-7.7$, consistent with the previous results, and seem to much better reproduce the observed photometry (reduced $\chi^2=5.65$ and $\chi^2=4.47$ for the \texttt{BEAGLE-AGN} and \texttt{CIGALE} fits, respectively) than the galaxy-only fits presented in section~\ref{sec:galaxy-SEDs}. The two AGN fits also yield lower dust attenuations than the galaxy-only fits (see Fig.~\ref{fig:SED-galaxies}) which are now more consistent with the ALMA non-detection. 

Assuming a covering fraction of 10\% by the narrow line emitting gas, from the \texttt{BEAGLE-AGN} fit we find an accretion luminosity of $\log(L_{\mathrm{acc}}/\mathrm{erg}\,\mathrm{s}^{-1})=44.86\pm0.03$. The \texttt{CIGALE} fit yields a lower accretion luminosity of $\log(L_{\mathrm{acc}}/\mathrm{erg}\,\mathrm{s}^{-1})=42.77\pm0.06$. Integrating the entire best-fitting AGN SEDs over their full wavelength range (i.e. from rest-frame $0.005\,\mu\mathrm{m}$ to $255\,\mu\mathrm{m}$ for \texttt{BEAGLE-AGN} and from rest-frame $0.001\,\mu\mathrm{m}$ to $1\times10^6\,\mu\mathrm{m}$ for \texttt{CIGALE}), we find bolometric luminosities of $L_{\mathrm{bol}}\simeq10^{44}\,\frac{\mathrm{erg}}{\mathrm{s}}$ for both the type~I and type~II AGN, complementing the measurement in section~\ref{sec:luminosity}.

The solution from \texttt{CIGALE}, which suggests a type~I AGN component, provides officially the better fit to the photometry among the two. In this fit, the UV continuum is dominated by the AGN component and the optical colors are produced by a very steep red stellar continuum (right-hand panel of Fig.~\ref{fig:SED-AGN}). This solution again requires a very large stellar mass of $\log(M_{\star}/\mathrm{M}_{\odot})=9.62\pm0.04$. In comparison, the \texttt{BEAGLE-AGN} fit yields a lower stellar mass ($\log(M_{\star}/\mathrm{M}_{\odot})=8.1_{-0.3}^{+0.2}$), and instead reproduces the red colors in the LW filters with extremely strong nebular emission powered by the type~II AGN component ($\mathrm{EW}_{[\mathrm{O\,III}]+\mathrm{H}\beta}=9137_{-1320}^{+1459}$\,\AA). In that respect one should note that most rest-frame UV emission lines predicted by the latter fit (black curve in the left-hand panel of Fig.~\ref{fig:SED-AGN}), are likely too weak -- i.e. with magnified integrated fluxes $F<1\times10^{-18}\,\frac{\mathrm{erg}}{\mathrm{s}\,\mathrm{cm}^2}$ -- to be detected in the GLASS-JWST NIRISS grism observations (see section~\ref{sec:data} and \citealt{treu22}) according to the JWST/NIRISS ETC. Hence, their non-detection does not currently help to discriminate between the two cases.

It is also important to note that in the presented SED fits, the extremely strong rest-frame optical [\ion{O}{3}] and H$\beta$ lines fall exactly onto the edge of the F410M and F444W bands which is also the reason for the seemingly very precise photometric redshift estimate (see Figs.~\ref{fig:SED-galaxies} and~\ref{fig:SED-AGN}). This is due to the fact that the LW bands have a much higher signal-to-noise than the other filters (see Tab.~\ref{tab:photometry}) and thus dominate the fit. We therefore run another set of \texttt{BEAGLE} tests assuming a minimum uncertainty of 5\% in each band. We do not show these explicitly here but note that the resulting fits, in particular the \texttt{BEAGLE-AGN} fit, are only marginally different. While in this 5\% minimum-error test the galaxy-only fit (see section~\ref{sec:galaxy-SEDs}) technically has a slightly lower $\chi^2$ than the \texttt{BEAGLE-AGN} fit shown here, it also predicts an extremely large stellar mass of $\sim10^{10}\,\mathrm{M}_{\odot}$, similar to but somewhat higher than the mass implied by the \texttt{CIGALE} fit, and a dust attenuation of $A_V\simeq5$. These are however disfavored due to the considerations of mass density and the ALMA non detection as already discussed in section~\ref{sec:galaxy-SEDs}.

\subsubsection{AGN-only SED fit} \label{sec:Ivo-SEDs}

\begin{figure}
    \centering
    \includegraphics[width=\columnwidth]{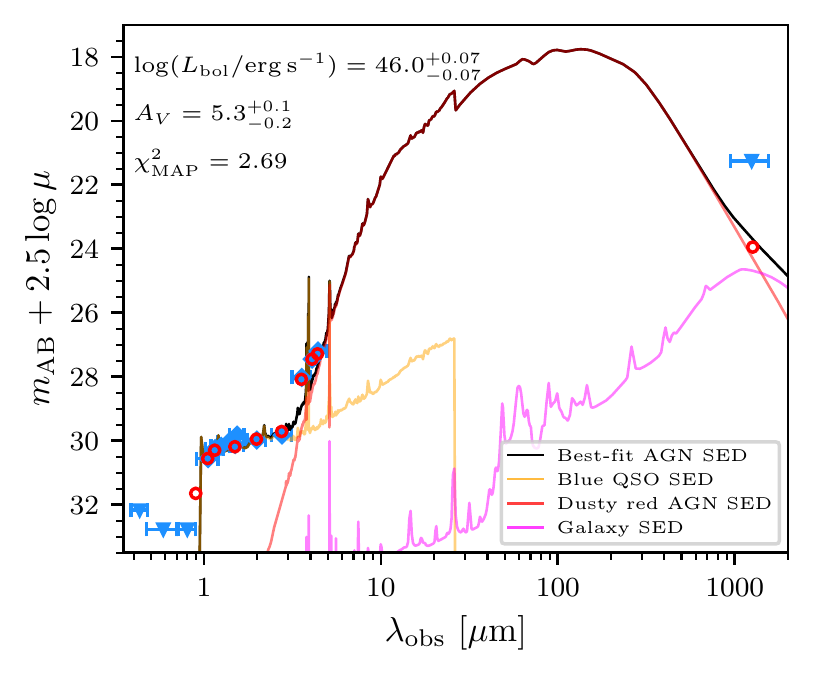}
    \caption{Maximum-a-posteriori (MAP) obscured AGN SED (black). As in Figs.~\ref{fig:SED-galaxies} and~\ref{fig:SED-AGN}, the measured and marginal photometries are shown in blue and red respectively. We decompose the best-fit SED into its reddened QSO component (red) and the scattered light that makes up the rest-frame UV component of the SED (orange). The contribution of the host galaxy is marginal and shown in purple. Since in this scenario the dust is warmer than in a typical dusty star-forming galaxy (cf. Fig.~\ref{fig:SED-galaxies}), its emission peaks at lower wavelengths than the $\lambda\sim1.2$\,mm probed by our ALMA Band~6 measurement such that the substantial dust attenuation necessary to fit the LW bands is nonetheless consistent with the ALMA non-detection.}
    \label{fig:SED-Ivo}
\end{figure}

As shown in the previous sections (\ref{sec:galaxy-SEDs} and \ref{sec:AGN-SEDs}), the combined galaxy and AGN SED fits suggest that the rest-frame UV part of the SED is best accounted for by a type~I AGN, and the red optical part is governed by a contribution from either emission lines or stellar continuum. However, none of these provides a very-likely fit to the object's photometry: the red optical part requires an underlying stellar component that is too massive given the object's size and also predicts an ALMA detection that is not observed, and the extreme emission-line solution requires significant contribution from nebular continuum which results in a much redder UV slope than observed. This might suggest that the object could in fact be AGN-dominated in both the rest-frame UV and optical regimes, which could be the case if we were looking at a heavily dust-obscured type~I AGN where the red continuum stems from the AGN-heated dust, the relatively faint blue continuum comes from scattered UV light, and the host galaxy does not contribute significantly at all \citep[e.g.][]{Glikman2013,Glikman2023,Banerji2015,Assef2016}.

We therefore perform an additional SED fit using a compound model based on stacked \textit{Sloan Digital Sky Survey} \citep[SDSS; e.g.][]{Stoughton2022} type~I QSOs \citep{VandenBerk2001} complemented with the \citet{Glikman2006AGNtemplate} template at longer wavelengths. The model has two components: a reddened component that represents a moderately dust-obscured ($A_V\sim3-5$) type~I AGN, and a non-reddened component that represents the scattered light of the QSO (typically $<1$\,\% of the red component's bolometric luminosity). The reddened component is generated by applying dust attenuation to the type~1 QSO template above, integrating the absorbed flux and re-emitting it in the IR using the AGN dust emission prescription implemented in \texttt{Prospector} which is described in detail in \citet{Leja2018ProspectorAGN}. This model was inspired by the NIRCam and NIRSpec observations of a similar yet less extreme, i.e. bluer and low-redshift, source in the CEERS program \citep{Kocevski2023} and manages to reproduce its emission lines very well (see Labb\'{e} et al. in prep.). The primary fit parameters of this simple obscured AGN model are the bolometric luminosity and the dust attenuation.

The best-fitting SED with this model is shown in Fig.~\ref{fig:SED-Ivo}. It manages to reproduce at the same time both the rest-frame UV and optical photometry of the source. As expected, the host galaxy does not contribute to the total photometry, meaning that the object is entirely dominated by the AGN. In addition, and unlike the galaxy-only SEDs presented in section~\ref{sec:galaxy-SEDs}, this scenario is consistent with the compact nature of our object (see section~\ref{sec:size}), its low UV luminosity (see section~\ref{sec:luminosity}) and the ALMA non-detection. The latter is because the dust would be warmer and its continuum therefore peak at lower wavelengths than the $\lambda\sim1.2$\,mm observed in ALMA Band~6 as can be seen in Fig.~\ref{fig:SED-Ivo}. The best-fitting SED therefore has a relatively high dust attenuation of $A_V=5.3_{-0.2}^{+0.1}$ and the resulting bolometric luminosity of $L_{\mathrm{bol}}\simeq10^{46}\,\frac{\mathrm{erg}}{\mathrm{s}}$ is higher by two orders of magnitude than in the fits presented in section~\ref{sec:AGN-SEDs}. 

It should be acknowledged that, since here we do not probe specifically such a combination, at this point it is unclear if we could differentiate between the above scenario, in which the red optical part comes from the reddened type~I QSO and the blue UV part comes from scattered, un-reddened UV light of the QSO, and a scenario in which the blue light comes from a low-mass un-obscured star-formation as in \citet{Akins2023}. We defer further SED analysis of this object to future work.

This two-component AGN-only fit is the best among all different SED fits tried in this work and we thus adopt it as the most probable scenario. We therefore conclude that this triply-imaged source is most likely a highly dust obscured red quasar.

\section{Discussion} \label{sec:discussion}
Given the SED-fitting results and the other properties of this object described in section~\ref{sec:results}, we tentatively conclude that this triply imaged, red compact object is most probably a UV-faint dust-obscured AGN at high redshift as presented in section~\ref{sec:Ivo-SEDs}. While throughout this study our primary candidate engine for the underlying emission of our object is a UV-faint dust-obscured AGN, or an intense and possibly dust-obscured star-forming clump -- which albeit less likely should perhaps not be completely ruled out -- we also briefly discuss in this section other possible candidates.

The first generations of stars to have formed in the Universe, i.e. \textit{Pop.~III stars}, are expected to have been very poor in metals, massive ($>10$\,M$_{\odot}$), energetic, and short lived \citep[e.g.][and references therein]{Zackrisson2011PopIII} and may have contributed significantly to the cosmic reionization. The red colors that we measure (see section~\ref{sec:color}), which could also be indicative of strong nebular emission lines as mentioned several times in this work, are however in disagreement with the colors expected for Pop.~III galaxies \citep[see][]{Zackrisson2011PopIII}. In addition, Pop.~III stars are expected to show strong Balmer lines but only weak lines of heavier elements such as [\ion{O}{3}], which would be in tension with the extremely high [\ion{O}{3}]+H$\beta$ EWs suggested by some of our SED-fitting analysis (section~\ref{sec:sed}). Note however that the latter does not contain Pop.~III templates and would have difficulties fitting such a population as discussed in detail in e.g. \citet{furtak22b}. The UV continuum slope expected for Pop.~III galaxies is also significantly bluer than our measurements (section~\ref{sec:color}), and we thus conclude that the emission of our object is likely not of Pop.~III origin.

\emph{Supermassive stars}, with masses above $10^3\,\mathrm{M}_{\odot}$, are a population of stars put forward to explain chemical abundances in globular clusters \citep[e.g.][and references therein] {Gieles2018MNRAS.478.2461G,Martins2020SupermassiveStars}. Given that they should reside in star clusters and are expected to have a non-typical spectral shape compared to `regular' stars, such as strong Balmer emission lines (as well as other emission or absorption properties that depend on temperature), we briefly consider them here as well. Indeed, the colors measured for the red object could possibly fit the predicted colors of supermassive stars (for example, see the (F115W-F200W) vs. (F200W-F444W) colors shown in Fig.~15 in \citealt{Martins2020SupermassiveStars}). Supermassive stars should therefore possibly be considered as a viable candidate population that is not ruled out currently.

\textit{Dark stars} are hypothesized stellar objects that are powered by DM annihilation, rather than atomic fusion \citep{Freese2016RPPh...79f6902F}. They are formed at about 1\,M$_{\odot}$ but can grow through accretion to large masses above $>10^{6}$\,M$_{\odot}$ and luminosities $>10^{10}$\,L$_{\odot}$, essentially becoming (some sort of) supermassive stars, such that they -- if they exist -- could be detected with the JWST. However, we find that the red colors we measure for our object are substantially redder than what is expected for these hypothetical dark stars (see Fig.~6 in \citealt{Freese2016RPPh...79f6902F}). 

A \emph{Pop.~III hypernova} is an ultra-energetic supernova of a Pop.~III star, typically with a mass of few M$_{\odot}$ and peak bolometric luminosity of a several 10$^{46}$\,L$_{\odot}$ \citep{Smidt2014ApJ...797...97S}. As these are expected in the early universe (every few years in an early bright galaxy \citep[e.g.][]{PadmanabhanLoeb2022GReGr..54...24P}) and are bright enough, they are worth considering as well. However, the time delays from our lens model (section~\ref{sec:SL}) suggest that the arrival time of QSO1C is about $\sim$19\,yr prior to QSO1A, and QSO1A arrives about $\sim3$\,yr prior to QSO1B, which is last to arrive. The total time delay between images C and B is thus about 22 years. Since the light-curves of supernovae, including hypernovae, only last several months \citep[e.g.][]{Smidt2014ApJ...797...97S}, we can exclude this as a likely option even at $z\sim8$. In general, any transient event with a source time-span smaller than 22\,yr divided by $(1+z)$ can be excluded.

\emph{DCBHs} are hypothesized black holes that are formed via a direct collapse of a primordial gas halo, rather than via stellar remnants, leading to black holes of about 10$^{5}$\,M$_{\odot}$ \citep[e.g.][]{Ferrara2014MNRAS.443.2410F,BrommLoeb2003ApJ...596...34B,Begelman2006MNRAS.370..289BDCBH}. DCBHs are of particular interest here because they are expected to have a redder SED than typical AGN \citep[e.g.][]{Pacucci2016DCBH}. Interestingly, our red object does indeed agree with the colors expected from DCBHs, suggesting in that framework a black-hole seed mass of nearly 10$^{5}$\,M$_{\odot}$ (see Fig.~4 in \citealt{Pacucci2016DCBH}; based on the F444W-$H_{\mathrm{F160W}}$ and F356W-$H_{\mathrm{F160W}}$ colors which we adopt as approximate alternatives to the $\sim4.5\,\mu\mathrm{m}$-$H_{\mathrm{F160W}}$ and $\sim3.6\,\mu\mathrm{m}$-$H_{\mathrm{F160W}}$ colors shown therein). 

Out of these more exotic objects considered above it seems that, based on the colors and on time delay arguments, only supermassive stars and DCBHs remain viable options in addition to the more common interpretations of an AGN or a red or extreme emission line clump (or very compact galaxy) which were extensively discussed in section~\ref{sec:results}. While indeed intriguing, our goal here is mainly to present the object and, while awaiting future spectra, we defer a more detailed analysis of these more exotic options to future work. A further investigation of this object will be possible once the UNCOVER NIRSpec observations become available by July 2023. Given its relative brightness in the LW bands ($m_{\mathrm{F444W}}\sim25-26$\,AB; see Tab.~\ref{tab:photometry}), our object should be easily detectable with the planned UNCOVER JWST/NIRSpec observations \citep[for setup details see][]{Bezanson2022arXiv221204026B}. These observations will enable us to disentangle the contributions of rest-frame optical continuum and emission lines and thus allow us to infer the true nature of the source. Note that if confirmed with spectroscopy, this would make this object comparable in distance to the highest-redshift AGN or candidates known to date \citep[e.g.][]{Wang_2021,Larson2023} and add to the small number of AGN that are known to be multiply imaged by galaxy clusters \citep[e.g.][]{Inada2003Natur.426..810I,Oguri2013MNRAS.429..482O,sharon2017,Acebron2022QSO,Acebron2022QSO2}. It will also constitute one of the UV-faintest AGN-dominated objects ever observed.

It should also be noted that the object is situated at a similar redshift as the well-known high-redshift proto-cluster behind A2744 \citep[e.g.][]{Zheng2014A2744} which was recently confirmed with JWST spectroscopy at $z=7.89$ \citep{Morishita2022arXiv221109097M}. De-projecting our object's position using our lens model, we find that it would lie only $\sim200$\,kpc away from the overdensity if it were at the same redshift which is possible given the known uncertainties and systematics of photometric redshift estimates.

\section{Conclusion}\label{sec:conclusion}
We present a unique, extremely red and compact object at $z_{\mathrm{phot}}\simeq7.6$ which is triply imaged by the SL galaxy cluster Abell~2744. The object was detected in recent deep multi-band JWST/NIRCam imaging taken for the UNCOVER program and its high-redshift nature is independently supported geometrically by the gravitational lensing (with a lower limit of $z_{\mathrm{geo}}\gtrsim5.5$).

Thanks to the lens magnification, we can limit the size of this object at $z\simeq7.6$ to $r_{e}\lesssim35$\,pc, suggesting that it is extremely compact. We measure for this object a rather red UV continuum slope of $\beta_{\mathrm{UV}}=-1.6\pm0.2$ and a relatively faint UV luminosity of $M_{\mathrm{UV},1450}=-16.81\pm0.09$, yet a high bolometric luminosity ($\gtrsim10^{43}\,\frac{\mathrm{erg}}{\mathrm{s}}$ and $\sim10^{44}-10^{46}\,\frac{\mathrm{erg}}{\mathrm{s}}$ respectively from the sum of observed fluxes and from the SED fits). Our object resides in significantly different locations on both color-color and $M_{\mathrm{UV}}$-size diagrams than the regions typically occupied by star-forming galaxies or the red high-redshift galaxy populations recently revealed with the JWST. Its compact size, UV slope, extremely red rest-frame optical colors, and luminosity density suggest that the emission might possibly be assisted -- or dominated -- by an AGN component. 

We use various codes to fit the photometry with different galaxy, galaxy+AGN, and AGN-only templates. From the galaxy-only fits, a dusty red galaxy SED matches the photometry reasonably well, although given the non-detection in dust continuum with ALMA band~6 this solution is somewhat disfavored. The addition of an AGN component improves the fit further and, while a type II AGN fit with extreme emission lines yields a reasonable fit as well, the best galaxy+AGN fit suggests a type I AGN that dominates the UV continuum, in combination with a massive, red galaxy component that dominates the rest-frame optical. These galaxy+AGN fits, however, do not seem to perfectly match the observed SED either. The best match to the data is finally provided by a two-component AGN-only fit, implying a heavily dust-obscured ($A_V=5.3_{-0.2}^{+0.1}$) AGN in which the rest-frame optical red continuum comes from reddened nuclear emission, i.e., warm dust emission powered by the AGN, and the bluer continuum in the rest-frame UV comes from un-obscured, scattered light from the AGN that escapes to the observer, or potentially -- low-mass un-obscured star formation. 

In addition to the AGN and galaxy interpretations, we also discuss other candidates such as e.g. a compact clump of Pop.~III, dark, or supermassive stars, as well as DCBHs. The predicted colors of Pop.~III or dark stars however do not seem to agree with the observed colors and strong emission lines needed to explain them, whereas the predicted color of supermassive star populations, and that of DCBHs, would agree with the observations and could therefore perhaps be considered an alternative scenario to the AGN. Spectroscopic observations planned for the next year with JWST/NIRSpec will be crucial for shedding light on the true nature of this unique source. If confirmed as an AGN at $z=7.6$, this object would be one of the most distant and faintest quasar-like objects observed to date, and the most distant known gravitationally lensed AGN.

%%%%%%%%%%%%%%%%%%%%%%%%%%%%%%%%%%%%%%%%%%%%%%%%%%%%
\acknowledgements

The authors would like to thank the anonymous referee for their very useful comments, which greatly helped to improve the paper. LF thanks Pierre Boldrini for very useful discussions about high-redshift star clusters. AZ thanks Erik Zackrisson for a useful discussion regarding Pop III stars. LF and AZ acknowledge support by grant 2020750 from the United States-Israel Binational Science Foundation (BSF) and grant 2109066 from the United States National Science Foundation (NSF), and by the Ministry of Science \& Technology, Israel. RB acknowledges support from the Research Corporation for Scientific Advancement (RCSA) Cottrell Scholar Award ID No.: 27587. ECL acknowledges support of an STFC Webb Fellowship (ST/W001438/1). HA acknowledges support from CNES (Centre National d'Etudes Spatiales). JC acknowledges funding from the ``FirstGalaxies'' Advanced Grant from the European Research Council (ERC) under the European Union’s Horizon 2020 research and innovation program (Grant agreement No.~789056). PD acknowledges support from the NWO grant~016.VIDI.189.162 (``ODIN") and from the European Commission's and University of Groningen's CO-FUND Rosalind Franklin program. The work of CCW is supported by NOIRLab, which is managed by the Association of Universities for Research in Astronomy (AURA) under a cooperative agreement with the National Science Foundation. The \textit{Cosmic Dawn Center} (DAWN) is funded by the Danish National Research Foundation (DNRF) under grant \#140. Support for the program JWST-GO-02561 was provided through a grant from the STScI under NASA contract NAS 5-03127. Cloud-based data processing and file storage for this work is provided by the AWS Cloud Credits for Research program.

This work is based on observations made with the NASA/ESA/CSA JWST and with the NASA/ESA \textit{Hubble Space Telescope} (HST). The data were obtained from the \texttt{Barbara A. Mikulski Archive for Space Telescopes} (\texttt{MAST}) at the \textit{Space Telescope Science Institute} (STScI), which is operated by the Association of Universities for Research in Astronomy (AURA), Inc., under NASA contract NAS 5-03127 for JWST. These observations are associated with the JWST GO program number 2561, JWST ERS program number 1324, JWST DD program number 2756 and HST GO programs 9722, 10493, 10793, and 12101. The specific observations used in this work can be accessed via \dataset[10.17909/nftp-e621]{http://dx.doi.org/10.17909/nftp-e621}. This paper also makes use of the following ALMA data: ADS/JAO.ALMA\#2018.1.00035.L and 2013.1.00999.S. ALMA is a partnership of ESO (representing its member states), NSF (USA) and NINS (Japan), together with NRC (Canada), MOST and ASIAA (Taiwan), and KASI (Republic of Korea), in cooperation with the Republic of Chile. The Joint ALMA Observatory is operated by ESO, AUI/NRAO and NAOJ. The National Radio Astronomy Observatory is a facility of the National Science Foundation operated under cooperative agreement by Associated Universities, Inc. This research has also made use of data obtained from the \textit{Chandra} Data Archive.

This research made use of \texttt{Astropy},\footnote{\url{http://www.astropy.org}} a community-developed core \texttt{python} package for Astronomy \citep{astropy13,astropy18} as well as the packages \texttt{NumPy} \citep{vanderwalt11}, \texttt{SciPy} \citep{virtanen20}, \texttt{Matplotlib} \citep{hunter07} and the \texttt{MAAT} Astronomy and Astrophysics tools for \texttt{MATLAB} \citep[][]{maat14}.

%%%%%%%%%%%%%%%%%%%%%%%%%%%%%%%%%%%%%%%%%%%%%%%%%%%%

%\clearpage
\newpage
\bibliographystyle{aasjournal} 
\bibliography{MyBiblio}

\begin{thebibliography}{}
\expandafter\ifx\csname natexlab\endcsname\relax\def\natexlab#1{#1}\fi
\providecommand{\url}[1]{\href{#1}{#1}}

\bibitem[{{Acebron} {et~al.}(2022{\natexlab{a}}){Acebron}, {Grillo},
  {Bergamini}, {Caminha}, {Tozzi}, {Mercurio}, {Rosati}, {Brammer},
  {Meneghetti}, {Nonino}, \& {Vanzella}}]{Acebron2022QSO}
{Acebron}, A., {Grillo}, C., {Bergamini}, P., {et~al.} 2022{\natexlab{a}},
  \aap, 668, A142

\bibitem[{{Acebron} {et~al.}(2022{\natexlab{b}}){Acebron}, {Grillo},
  {Bergamini}, {Mercurio}, {Rosati}, {Caminha}, {Tozzi}, {Brammer},
  {Meneghetti}, {Morelli}, {Nonino}, \& {Vanzella}}]{Acebron2022QSO2}
---. 2022{\natexlab{b}}, \apj, 926, 86

\bibitem[{{Adams} {et~al.}(2023){Adams}, {Conselice}, {Ferreira}, {Austin},
  {Trussler}, {Juod{\v{z}}balis}, {Wilkins}, {Caruana}, {Dayal}, {Verma}, \&
  {Vijayan}}]{Adams2023MNRAS.518.4755A}
{Adams}, N.~J., {Conselice}, C.~J., {Ferreira}, L., {et~al.} 2023, \mnras, 518,
  4755

\bibitem[{{Akins} {et~al.}(2023){Akins}, {Casey}, {Allen}, {Bagley},
  {Dickinson}, {Finkelstein}, {Franco}, {Harish}, {Arrabal Haro}, {Ilbert},
  {Kartaltepe}, {Koekemoer}, {Liu}, {Long}, {McCracken}, {Paquereau},
  {Papovich}, {Pirzkal}, {Rhodes}, {Robertson}, {Shuntov}, {Toft}, {Yang},
  {Barro}, {Bisigello}, {Buat}, {Champagne}, {Cooper}, {Costantin}, {de la
  Vega}, {Drakos}, {Faisst}, {Fontana}, {Fujimoto}, {Gillman},
  {G{\'o}mez-Guijarro}, {Gozaliasl}, {Hathi}, {Hayward}, {Hirschmann},
  {Holwerda}, {Jin}, {Kocevski}, {Kokorev}, {Lambrides}, {Lucas}, {Magdis},
  {Magnelli}, {McKinney}, {Mobasher}, {P{\'e}rez-Gonz{\'a}lez}, {Rich},
  {Seill{\'e}}, {Talia}, {Urry}, {Valentino}, {Whitaker}, {Yung}, \&
  {Zavala}}]{Akins2023}
{Akins}, H.~B., {Casey}, C.~M., {Allen}, N., {et~al.} 2023, arXiv e-prints,
  arXiv:2304.12347

\bibitem[{{Assef} {et~al.}(2016){Assef}, {Walton}, {Brightman}, {Stern},
  {Alexander}, {Bauer}, {Blain}, {Diaz-Santos}, {Eisenhardt}, {Finkelstein},
  {Hickox}, {Tsai}, \& {Wu}}]{Assef2016}
{Assef}, R.~J., {Walton}, D.~J., {Brightman}, M., {et~al.} 2016, \apj, 819, 111

\bibitem[{{Astropy Collaboration} {et~al.}(2013){Astropy Collaboration},
  {Robitaille}, {Tollerud}, {Greenfield}, {Droettboom}, {Bray}, {Aldcroft},
  {Davis}, {Ginsburg}, {Price-Whelan}, {Kerzendorf}, {Conley}, {Crighton},
  {Barbary}, {Muna}, {Ferguson}, {Grollier}, {Parikh}, {Nair}, {Unther},
  {Deil}, {Woillez}, {Conseil}, {Kramer}, {Turner}, {Singer}, {Fox}, {Weaver},
  {Zabalza}, {Edwards}, {Azalee Bostroem}, {Burke}, {Casey}, {Crawford},
  {Dencheva}, {Ely}, {Jenness}, {Labrie}, {Lim}, {Pierfederici}, {Pontzen},
  {Ptak}, {Refsdal}, {Servillat}, \& {Streicher}}]{astropy13}
{Astropy Collaboration}, {Robitaille}, T.~P., {Tollerud}, E.~J., {et~al.} 2013,
  \aap, 558, A33

\bibitem[{{Atek} {et~al.}(2018){Atek}, {Richard}, {Kneib}, \&
  {Schaerer}}]{atek18}
{Atek}, H., {Richard}, J., {Kneib}, J.-P., \& {Schaerer}, D. 2018, \mnras, 479,
  5184

\bibitem[{{Atek} {et~al.}(2014){Atek}, {Richard}, {Kneib}, {Clement}, {Egami},
  {Ebeling}, {Jauzac}, {Jullo}, {Laporte}, {Limousin}, \&
  {Natarajan}}]{Atek2014A2744}
{Atek}, H., {Richard}, J., {Kneib}, J.-P., {et~al.} 2014, \apj, 786, 60

\bibitem[{{Atek} {et~al.}(2015{\natexlab{a}}){Atek}, {Richard}, {Jauzac},
  {Kneib}, {Natarajan}, {Limousin}, {Schaerer}, {Jullo}, {Ebeling}, {Egami}, \&
  {Clement}}]{atek15b}
{Atek}, H., {Richard}, J., {Jauzac}, M., {et~al.} 2015{\natexlab{a}}, \apj,
  814, 69

\bibitem[{{Atek} {et~al.}(2015{\natexlab{b}}){Atek}, {Richard}, {Kneib},
  {Jauzac}, {Schaerer}, {Clement}, {Limousin}, {Jullo}, {Natarajan}, {Egami},
  \& {Ebeling}}]{atek15a}
{Atek}, H., {Richard}, J., {Kneib}, J.-P., {et~al.} 2015{\natexlab{b}}, \apj,
  800, 18

\bibitem[{{Atek} {et~al.}(2023){Atek}, {Shuntov}, {Furtak}, {Richard}, {Kneib},
  {Mahler}, {Zitrin}, {McCracken}, {Charlot}, {Chevallard}, \&
  {Chemerynska}}]{Atek2023}
{Atek}, H., {Shuntov}, M., {Furtak}, L.~J., {et~al.} 2023, \mnras, 519, 1201

\bibitem[{{Ba{\~n}ados} {et~al.}(2018){Ba{\~n}ados}, {Carilli}, {Walter},
  {Momjian}, {Decarli}, {Farina}, {Mazzucchelli}, \&
  {Venemans}}]{Banados2018z7p54Quasar}
{Ba{\~n}ados}, E., {Carilli}, C., {Walter}, F., {et~al.} 2018, \apjl, 861, L14

\bibitem[{{Ba{\~n}ados} {et~al.}(2016){Ba{\~n}ados}, {Venemans}, {Decarli},
  {Farina}, {Mazzucchelli}, {Walter}, {Fan}, {Stern}, {Schlafly}, {Chambers},
  {Rix}, {Jiang}, {McGreer}, {Simcoe}, {Wang}, {Yang}, {Morganson}, {De Rosa},
  {Greiner}, {Balokovi{\'c}}, {Burgett}, {Cooper}, {Draper}, {Flewelling},
  {Hodapp}, {Jun}, {Kaiser}, {Kudritzki}, {Magnier}, {Metcalfe}, {Miller},
  {Schindler}, {Tonry}, {Wainscoat}, {Waters}, \&
  {Yang}}]{Banados2016100PS1Quasars}
{Ba{\~n}ados}, E., {Venemans}, B.~P., {Decarli}, R., {et~al.} 2016, \apjs, 227,
  11

\bibitem[{{Bacon} {et~al.}(2010){Bacon}, {Accardo}, {Adjali}, {Anwand},
  {Bauer}, {Biswas}, {Blaizot}, {Boudon}, {Brau-Nogue}, {Brinchmann},
  {Caillier}, {Capoani}, {Carollo}, {Contini}, {Couderc}, {Daguis{\'e}},
  {Deiries}, {Delabre}, {Dreizler}, {Dubois}, {Dupieux}, {Dupuy}, {Emsellem},
  {Fechner}, {Fleischmann}, {Fran{\c{c}}ois}, {Gallou}, {Gharsa}, {Glindemann},
  {Gojak}, {Guiderdoni}, {Hansali}, {Hahn}, {Jarno}, {Kelz}, {Koehler},
  {Kosmalski}, {Laurent}, {Le Floch}, {Lilly}, {Lizon}, {Loupias}, {Manescau},
  {Monstein}, {Nicklas}, {Olaya}, {Pares}, {Pasquini}, {P{\'e}contal-Rousset},
  {Pell{\'o}}, {Petit}, {Popow}, {Reiss}, {Remillieux}, {Renault}, {Roth},
  {Rupprecht}, {Serre}, {Schaye}, {Soucail}, {Steinmetz}, {Streicher}, {Stuik},
  {Valentin}, {Vernet}, {Weilbacher}, {Wisotzki}, \& {Yerle}}]{Bacon2010MUSE}
{Bacon}, R., {Accardo}, M., {Adjali}, L., {et~al.} 2010, in Society of
  Photo-Optical Instrumentation Engineers (SPIE) Conference Series, Vol. 7735,
  Ground-based and Airborne Instrumentation for Astronomy III, ed. I.~S.
  {McLean}, S.~K. {Ramsay}, \& H.~{Takami}, 773508

\bibitem[{{Bagley} {et~al.}(2022){Bagley}, {Finkelstein}, {Koekemoer},
  {Ferguson}, {Arrabal Haro}, {Dickinson}, {Kartaltepe}, {Papovich},
  {P{\'e}rez-Gonz{\'a}lez}, {Pirzkal}, {Somerville}, {Willmer}, {Yang}, {Yung},
  {Fontana}, {Grazian}, {Grogin}, {Hirschmann}, {Kewley}, {Kirkpatrick},
  {Kocevski}, {Lotz}, {Medrano}, {Morales}, {Pentericci}, {Ravindranath},
  {Trump}, {Wilkins}, {Calabr{\`o}}, {Cooper}, {Costantin}, {de la Vega},
  {Hutchison}, {Lucas}, {McGrath}, {Wang}, \& {Wuyts}}]{bagley2022}
{Bagley}, M.~B., {Finkelstein}, S.~L., {Koekemoer}, A.~M., {et~al.} 2022, arXiv
  e-prints, arXiv:2211.02495

\bibitem[{{Banerji} {et~al.}(2015){Banerji}, {Alaghband-Zadeh}, {Hewett}, \&
  {McMahon}}]{Banerji2015}
{Banerji}, M., {Alaghband-Zadeh}, S., {Hewett}, P.~C., \& {McMahon}, R.~G.
  2015, \mnras, 447, 3368

\bibitem[{Barbary(2016)}]{barbary16}
Barbary, K. 2016, Journal of Open Source Software, 1, 58.
\newblock \url{https://doi.org/10.21105/joss.00058}

\bibitem[{{Barbary}(2018)}]{barabry18}
{Barbary}, K. 2018, {SEP: Source Extraction and Photometry}, Astrophysics
  Source Code Library, record ascl:1811.004, , , ascl:1811.004

\bibitem[{{Barro} {et~al.}(2014){Barro}, {Faber}, {P{\'e}rez-Gonz{\'a}lez},
  {Pacifici}, {Trump}, {Koo}, {Wuyts}, {Guo}, {Bell}, {Dekel}, {Porter},
  {Primack}, {Ferguson}, {Ashby}, {Caputi}, {Ceverino}, {Croton}, {Fazio},
  {Giavalisco}, {Hsu}, {Kocevski}, {Koekemoer}, {Kurczynski}, {Kollipara},
  {Lee}, {McIntosh}, {McGrath}, {Moody}, {Somerville}, {Papovich}, {Salvato},
  {Santini}, {Tal}, {van der Wel}, {Williams}, {Willner}, \&
  {Zolotov}}]{barro2014}
{Barro}, G., {Faber}, S.~M., {P{\'e}rez-Gonz{\'a}lez}, P.~G., {et~al.} 2014,
  \apj, 791, 52

\bibitem[{{Beelen} {et~al.}(2006){Beelen}, {Cox}, {Benford}, {Dowell},
  {Kov{\'a}cs}, {Bertoldi}, {Omont}, \& {Carilli}}]{Beelen2006}
{Beelen}, A., {Cox}, P., {Benford}, D.~J., {et~al.} 2006, \apj, 642, 694

\bibitem[{{Begelman} {et~al.}(2006){Begelman}, {Volonteri}, \&
  {Rees}}]{Begelman2006MNRAS.370..289BDCBH}
{Begelman}, M.~C., {Volonteri}, M., \& {Rees}, M.~J. 2006, \mnras, 370, 289

\bibitem[{{Bergamini} {et~al.}(2023){Bergamini}, {Acebron}, {Grillo}, {Rosati},
  {Caminha}, {Mercurio}, {Vanzella}, {Angora}, {Brammer}, {Meneghetti}, \&
  {Nonino}}]{bergamini22}
{Bergamini}, P., {Acebron}, A., {Grillo}, C., {et~al.} 2023, \aap, 670, A60

\bibitem[{{Bertin} \& {Arnouts}(1996)}]{BertinArnouts1996Sextractor}
{Bertin}, E., \& {Arnouts}, S. 1996, \aaps, 117, 393

\bibitem[{{Bezanson} {et~al.}(2022){Bezanson}, {Labbe}, {Whitaker}, {Leja},
  {Price}, {Franx}, {Brammer}, {Marchesini}, {Zitrin}, {Wang}, {Weaver},
  {Furtak}, {Atek}, {Coe}, {Cutler}, {Dayal}, {van Dokkum}, {Feldmann},
  {Forster Schreiber}, {Fujimoto}, {Geha}, {Glazebrook}, {de Graaff}, {Juneau},
  {Kassin}, {Kriek}, {Khullar}, {Greene}, {Maseda}, {Oesch}, {Smit},
  {Stefanon}, {Taylor}, \& {Williams}}]{Bezanson2022arXiv221204026B}
{Bezanson}, R., {Labbe}, I., {Whitaker}, K.~E., {et~al.} 2022, arXiv e-prints,
  arXiv:2212.04026

\bibitem[{{Bogdan} {et~al.}(2023){Bogdan}, {Goulding}, {Natarajan}, {Kovacs},
  {Tremblay}, {Chadayammuri}, {Volonteri}, {Kraft}, {Forman}, {Jones},
  {Churazov}, \& {Zhuravleva}}]{Bogdan2023}
{Bogdan}, A., {Goulding}, A., {Natarajan}, P., {et~al.} 2023, arXiv e-prints,
  arXiv:2305.15458

\bibitem[{{B{\"o}ker} {et~al.}(2023){B{\"o}ker}, {Beck}, {Birkmann},
  {Giardino}, {Keyes}, {Kumari}, {Muzerolle}, {Rawle}, {Zeidler}, {Abul-Huda},
  {Alves de Oliveira}, {Arribas}, {Bechtold}, {Bhatawdekar}, {Bonaventura},
  {Bunker}, {Cameron}, {Carniani}, {Charlot}, {Curti}, {Espinoza}, {Ferruit},
  {Franx}, {Jakobsen}, {Karakla}, {L{\'o}pez-Caniego}, {L{\"u}tzgendorf},
  {Maiolino}, {Manjavacas}, {Marston}, {Moseley}, {Ogle}, {Perna},
  {Pe{\~n}a-Guerrero}, {Pirzkal}, {Plesha}, {Proffitt}, {Rauscher}, {Rix},
  {Rodr{\'\i}guez del Pino}, {Rustamkulov}, {Sabbi}, {Sing}, {Sirianni}, {te
  Plate}, {{\'U}beda}, {Wahlgren}, {Wislowski}, {Wu}, \&
  {Willott}}]{Boeker2023}
{B{\"o}ker}, T., {Beck}, T.~L., {Birkmann}, S.~M., {et~al.} 2023, \pasp, 135,
  038001

\bibitem[{{Boquien} {et~al.}(2019){Boquien}, {Burgarella}, {Roehlly}, {Buat},
  {Ciesla}, {Corre}, {Inoue}, \& {Salas}}]{Boquien2019}
{Boquien}, M., {Burgarella}, D., {Roehlly}, Y., {et~al.} 2019, \aap, 622, A103

\bibitem[{{Bosman} {et~al.}(2022){Bosman}, {Davies}, {Becker}, {Keating},
  {Davies}, {Zhu}, {Eilers}, {D'Odorico}, {Bian}, {Bischetti}, {Cristiani},
  {Fan}, {Farina}, {Haehnelt}, {Hennawi}, {Kulkarni}, {Mesinger}, {Meyer},
  {Onoue}, {Pallottini}, {Qin}, {Ryan-Weber}, {Schindler}, {Walter}, {Wang}, \&
  {Yang}}]{Bosman2022MNRAS.514...55B}
{Bosman}, S. E.~I., {Davies}, F.~B., {Becker}, G.~D., {et~al.} 2022, \mnras,
  514, 55

\bibitem[{{Bouwens} {et~al.}(2023){Bouwens}, {Illingworth}, {Oesch},
  {Stefanon}, {Naidu}, {van Leeuwen}, \& {Magee}}]{Bouwens2022LFJWST}
{Bouwens}, R., {Illingworth}, G., {Oesch}, P., {et~al.} 2023, \mnras, 523, 1009

\bibitem[{{Bouwens} {et~al.}(2022){Bouwens}, {Illingworth}, {Ellis}, {Oesch},
  \& {Stefanon}}]{bouwens2022b}
{Bouwens}, R.~J., {Illingworth}, G., {Ellis}, R.~S., {Oesch}, P., \&
  {Stefanon}, M. 2022, \apj, 940, 55

\bibitem[{{Bouwens} {et~al.}(2017){Bouwens}, {Oesch}, {Illingworth}, {Ellis},
  \& {Stefanon}}]{Bouwens2017}
{Bouwens}, R.~J., {Oesch}, P.~A., {Illingworth}, G.~D., {Ellis}, R.~S., \&
  {Stefanon}, M. 2017, \apj, 843, 129

\bibitem[{{Bouwens} {et~al.}(2014){Bouwens}, {Illingworth}, {Oesch},
  {Labb{\'e}}, {van Dokkum}, {Trenti}, {Franx}, {Smit}, {Gonzalez}, \&
  {Magee}}]{Bouwens2014UVslopes}
{Bouwens}, R.~J., {Illingworth}, G.~D., {Oesch}, P.~A., {et~al.} 2014, \apj,
  793, 115

\bibitem[{{Bowler} {et~al.}(2017){Bowler}, {Dunlop}, {McLure}, \&
  {McLeod}}]{bowler2017}
{Bowler}, R.~A.~A., {Dunlop}, J.~S., {McLure}, R.~J., \& {McLeod}, D.~J. 2017,
  \mnras, 466, 3612

\bibitem[{Bradley {et~al.}(2022)Bradley, Sipőcz, Robitaille, Tollerud,
  Vinícius, Deil, Barbary, Wilson, Busko, Donath, Günther, Cara, Lim,
  Meßlinger, Conseil, Bostroem, Droettboom, Bray, Bratholm, Barentsen, Craig,
  Ginsburg, Rathi, Pascual, Perren, Georgiev, de~Val-Borro, Kerzendorf, Bach,
  \& Quint}]{photutils22}
Bradley, L., Sipőcz, B., Robitaille, T., {et~al.} 2022, astropy/photutils:
  1.6.0, v1.6.0,  Zenodo, doi:10.5281/zenodo.7419741.
\newblock \url{https://doi.org/10.5281/zenodo.7419741}

\bibitem[{Brammer {et~al.}(2022)Brammer, Strait, Matharu, \&
  Momcheva}]{grizli2022}
Brammer, G., Strait, V., Matharu, J., \& Momcheva, I. 2022, grizli, v1.5.0,
  Zenodo, please cite this software using these metadata.,
  doi:10.5281/zenodo.6672538.
\newblock \url{https://doi.org/10.5281/zenodo.6672538}

\bibitem[{{Brammer} {et~al.}(2008){Brammer}, {van Dokkum}, \&
  {Coppi}}]{Brammer2008EAZY}
{Brammer}, G.~B., {van Dokkum}, P.~G., \& {Coppi}, P. 2008, \apj, 686, 1503

\bibitem[{{Bromm} \& {Loeb}(2003)}]{BrommLoeb2003ApJ...596...34B}
{Bromm}, V., \& {Loeb}, A. 2003, \apj, 596, 34

\bibitem[{{Bruzual} \& {Charlot}(2003)}]{BC03}
{Bruzual}, G., \& {Charlot}, S. 2003, \mnras, 344, 1000

\bibitem[{{Calzetti} {et~al.}(2000){Calzetti}, {Armus}, {Bohlin}, {Kinney},
  {Koornneef}, \& {Storchi-Bergmann}}]{Calzetti2000}
{Calzetti}, D., {Armus}, L., {Bohlin}, R.~C., {et~al.} 2000, \apj, 533, 682

\bibitem[{{Carnall} {et~al.}(2023){Carnall}, {McLure}, {Dunlop}, {McLeod},
  {Wild}, {Cullen}, {Magee}, {Begley}, {Cimatti}, {Donnan}, {Hamadouche},
  {Jewell}, \& {Walker}}]{Carnall2023}
{Carnall}, A.~C., {McLure}, R.~J., {Dunlop}, J.~S., {et~al.} 2023, arXiv
  e-prints, arXiv:2301.11413

\bibitem[{{Castellano} {et~al.}(2023){Castellano}, {Fontana}, {Treu}, {Merlin},
  {Santini}, {Bergamini}, {Grillo}, {Rosati}, {Acebron}, {Leethochawalit},
  {Paris}, {Bonchi}, {Belfiori}, {Calabr{\`o}}, {Correnti}, {Nonino},
  {Polenta}, {Trenti}, {Boyett}, {Brammer}, {Broadhurst}, {Caminha}, {Chen},
  {Filippenko}, {Fortuni}, {Glazebrook}, {Mascia}, {Mason}, {Menci},
  {Meneghetti}, {Mercurio}, {Metha}, {Morishita}, {Nanayakkara}, {Pentericci},
  {Roberts-Borsani}, {Roy}, {Vanzella}, {Vulcani}, {Yang}, \&
  {Wang}}]{Castellano2022arXiv221206666C}
{Castellano}, M., {Fontana}, A., {Treu}, T., {et~al.} 2023, \apjl, 948, L14

\bibitem[{{Chabrier}(2003)}]{chabrier03}
{Chabrier}, G. 2003, \pasp, 115, 763

\bibitem[{{Charlot} \& {Fall}(2000)}]{CF2000}
{Charlot}, S., \& {Fall}, S.~M. 2000, \apj, 539, 718

\bibitem[{{Chevallard} \& {Charlot}(2016)}]{chevallard16}
{Chevallard}, J., \& {Charlot}, S. 2016, \mnras, 462, 1415

\bibitem[{{Coe} {et~al.}(2013){Coe}, {Zitrin}, {Carrasco}, {Shu}, {Zheng},
  {Postman}, {Bradley}, {Koekemoer}, {Bouwens}, {Broadhurst}, {Monna}, {Host},
  {Moustakas}, {Ford}, {Moustakas}, {van der Wel}, {Donahue}, {Rodney},
  {Ben{\'{\i}}tez}, {Jouvel}, {Seitz}, {Kelson}, \& {Rosati}}]{Coe2012highz}
{Coe}, D., {Zitrin}, A., {Carrasco}, M., {et~al.} 2013, \apj, 762, 32

\bibitem[{{Cowie} {et~al.}(1996){Cowie}, {Songaila}, {Hu}, \&
  {Cohen}}]{Cowie1996}
{Cowie}, L.~L., {Songaila}, A., {Hu}, E.~M., \& {Cohen}, J.~G. 1996, \aj, 112,
  839

\bibitem[{{Cui} {et~al.}(2021){Cui}, {Dav{\'e}}, {Peacock},
  {Angl{\'e}s-Alc{\'a}zar}, \& {Yang}}]{Cui2021NatAs...5.1069C}
{Cui}, W., {Dav{\'e}}, R., {Peacock}, J.~A., {Angl{\'e}s-Alc{\'a}zar}, D., \&
  {Yang}, X. 2021, Nature Astronomy, 5, 1069

\bibitem[{{Cullen} {et~al.}(2023){Cullen}, {McLure}, {McLeod}, {Dunlop},
  {Donnan}, {Carnall}, {Bowler}, {Begley}, {Hamadouche}, \&
  {Stanton}}]{Cullen2022UVslopes}
{Cullen}, F., {McLure}, R.~J., {McLeod}, D.~J., {et~al.} 2023, \mnras, 520, 14

\bibitem[{{Curtis-Lake} {et~al.}(2023){Curtis-Lake}, {Carniani}, {Cameron},
  {Charlot}, {Jakobsen}, {Maiolino}, {Bunker}, {Witstok}, {Smit}, {Chevallard},
  {Willott}, {Ferruit}, {Arribas}, {Bonaventura}, {Curti}, {D'Eugenio},
  {Franx}, {Giardino}, {Looser}, {L{\"u}tzgendorf}, {Maseda}, {Rawle}, {Rix},
  {Rodr{\'\i}guez del Pino}, {{\"U}bler}, {Sirianni}, {Dressler}, {Egami},
  {Eisenstein}, {Endsley}, {Hainline}, {Hausen}, {Johnson}, {Rieke},
  {Robertson}, {Shivaei}, {Stark}, {Tacchella}, {Williams}, {Willmer},
  {Bhatawdekar}, {Bowler}, {Boyett}, {Chen}, {de Graaff}, {Helton}, {Hviding},
  {Jones}, {Kumari}, {Lyu}, {Nelson}, {Perna}, {Sandles}, {Saxena}, {Suess},
  {Sun}, {Topping}, {Wallace}, \& {Whitler}}]{Curtis-Lake2022arXiv221204568C}
{Curtis-Lake}, E., {Carniani}, S., {Cameron}, A., {et~al.} 2023, Nature
  Astronomy, 7, 622

\bibitem[{{Donnan} {et~al.}(2023){Donnan}, {McLeod}, {Dunlop}, {McLure},
  {Carnall}, {Begley}, {Cullen}, {Hamadouche}, {Bowler}, {Magee}, {McCracken},
  {Milvang-Jensen}, {Moneti}, \& {Targett}}]{Donnan2022highz}
{Donnan}, C.~T., {McLeod}, D.~J., {Dunlop}, J.~S., {et~al.} 2023, \mnras, 518,
  6011

\bibitem[{{Doyon} {et~al.}(2012){Doyon}, {Hutchings}, {Beaulieu}, {Albert},
  {Lafreni{\`e}re}, {Willott}, {Touahri}, {Rowlands}, {Maszkiewicz},
  {Fullerton}, {Volk}, {Martel}, {Chayer}, {Sivaramakrishnan}, {Abraham},
  {Ferrarese}, {Jayawardhana}, {Johnstone}, {Meyer}, {Pipher}, \&
  {Sawicki}}]{doyon12}
{Doyon}, R., {Hutchings}, J.~B., {Beaulieu}, M., {et~al.} 2012, in Society of
  Photo-Optical Instrumentation Engineers (SPIE) Conference Series, Vol. 8442,
  Space Telescopes and Instrumentation 2012: Optical, Infrared, and Millimeter
  Wave, ed. M.~C. {Clampin}, G.~G. {Fazio}, H.~A. {MacEwen}, \& J.~{Oschmann},
  Jacobus~M., 84422R

\bibitem[{{Draine} {et~al.}(2014){Draine}, {Aniano}, {Krause}, {Groves},
  {Sandstrom}, {Braun}, {Leroy}, {Klaas}, {Linz}, {Rix}, {Schinnerer},
  {Schmiedeke}, \& {Walter}}]{Draine2014}
{Draine}, B.~T., {Aniano}, G., {Krause}, O., {et~al.} 2014, \apj, 780, 172

\bibitem[{{El{\'\i}asd{\'o}ttir} {et~al.}(2007){El{\'\i}asd{\'o}ttir},
  {Limousin}, {Richard}, {Hjorth}, {Kneib}, {Natarajan}, {Pedersen}, {Jullo},
  \& {Paraficz}}]{Eliasdottir2007arXiv0710.5636E}
{El{\'\i}asd{\'o}ttir}, {\'A}., {Limousin}, M., {Richard}, J., {et~al.} 2007,
  arXiv e-prints, arXiv:0710.5636

\bibitem[{{Endsley} {et~al.}(2022{\natexlab{a}}){Endsley}, {Stark}, {Whitler},
  {Topping}, {Chen}, {Plat}, {Chisholm}, \&
  {Charlot}}]{Endsley2022arXiv220814999E}
{Endsley}, R., {Stark}, D.~P., {Whitler}, L., {et~al.} 2022{\natexlab{a}},
  arXiv e-prints, arXiv:2208.14999

\bibitem[{{Endsley} {et~al.}(2022{\natexlab{b}}){Endsley}, {Stark}, {Fan},
  {Smit}, {Wang}, {Yang}, {Hainline}, {Lyu}, {Bouwens}, \&
  {Schouws}}]{Endsley2022AGNcandidate_MNRAS.512.4248E}
{Endsley}, R., {Stark}, D.~P., {Fan}, X., {et~al.} 2022{\natexlab{b}}, \mnras,
  512, 4248

\bibitem[{{Endsley} {et~al.}(2023){Endsley}, {Stark}, {Lyu}, {Wang}, {Yang},
  {Fan}, {Smit}, {Bouwens}, {Hainline}, \&
  {Schouws}}]{Endsley2022AGN_arXiv220600018E}
{Endsley}, R., {Stark}, D.~P., {Lyu}, J., {et~al.} 2023, \mnras, 520, 4609

\bibitem[{{Fan} {et~al.}(2006){Fan}, {Strauss}, {Becker}, {White}, {Gunn},
  {Knapp}, {Richards}, {Schneider}, {Brinkmann}, \&
  {Fukugita}}]{Fan2006Quasars}
{Fan}, X., {Strauss}, M.~A., {Becker}, R.~H., {et~al.} 2006, \aj, 132, 117

\bibitem[{{Fan} {et~al.}(2019){Fan}, {Barth}, {Banados}, {De Rosa}, {Decarli},
  {Eilers}, {Farina}, {Greene}, {Habouzit}, {Jiang}, {Jun}, {Koekemoer},
  {Malhotra}, {Mazzucchelli}, {Pacucci}, {Rhoads}, {Riechers}, {Rigby}, {Shen},
  {Simcoe}, {Stern}, {Strauss}, {Treu}, {Venemans}, {Vestergaard}, {Volonteri},
  {Walter}, {Yang}, \& {Wang}}]{Fan2019BAAS...51c.121F}
{Fan}, X., {Barth}, A., {Banados}, E., {et~al.} 2019, \baas, 51, 121

\bibitem[{{Feltre} {et~al.}(2016){Feltre}, {Charlot}, \& {Gutkin}}]{feltre2016}
{Feltre}, A., {Charlot}, S., \& {Gutkin}, J. 2016, \mnras, 456, 3354

\bibitem[{{Ferland} {et~al.}(2013){Ferland}, {Porter}, {van Hoof}, {Williams},
  {Abel}, {Lykins}, {Shaw}, {Henney}, \& {Stancil}}]{ferland13}
{Ferland}, G.~J., {Porter}, R.~L., {van Hoof}, P.~A.~M., {et~al.} 2013, \rmxaa,
  49, 137

\bibitem[{{Ferrara} {et~al.}(2014){Ferrara}, {Salvadori}, {Yue}, \&
  {Schleicher}}]{Ferrara2014MNRAS.443.2410F}
{Ferrara}, A., {Salvadori}, S., {Yue}, B., \& {Schleicher}, D. 2014, \mnras,
  443, 2410

\bibitem[{{Ferruit} {et~al.}(2022){Ferruit}, {Jakobsen}, {Giardino}, {Rawle},
  {Alves de Oliveira}, {Arribas}, {Beck}, {Birkmann}, {B{\"o}ker}, {Bunker},
  {Charlot}, {de Marchi}, {Franx}, {Henry}, {Karakla}, {Kassin}, {Kumari},
  {L{\'o}pez-Caniego}, {L{\"u}tzgendorf}, {Maiolino}, {Manjavacas}, {Marston},
  {Moseley}, {Muzerolle}, {Pirzkal}, {Rauscher}, {Rix}, {Sabbi}, {Sirianni},
  {te Plate}, {Valenti}, {Willott}, \& {Zeidler}}]{Ferruit2022}
{Ferruit}, P., {Jakobsen}, P., {Giardino}, G., {et~al.} 2022, \aap, 661, A81

\bibitem[{{Finkelstein} {et~al.}(2023){Finkelstein}, {Bagley}, {Ferguson},
  {Wilkins}, {Kartaltepe}, {Papovich}, {Yung}, {Haro}, {Behroozi}, {Dickinson},
  {Kocevski}, {Koekemoer}, {Larson}, {Le Bail}, {Morales},
  {P{\'e}rez-Gonz{\'a}lez}, {Burgarella}, {Dav{\'e}}, {Hirschmann},
  {Somerville}, {Wuyts}, {Bromm}, {Casey}, {Fontana}, {Fujimoto}, {Gardner},
  {Giavalisco}, {Grazian}, {Grogin}, {Hathi}, {Hutchison}, {Jha}, {Jogee},
  {Kewley}, {Kirkpatrick}, {Long}, {Lotz}, {Pentericci}, {Pierel}, {Pirzkal},
  {Ravindranath}, {Ryan}, {Trump}, {Yang}, {Bhatawdekar}, {Bisigello}, {Buat},
  {Calabr{\`o}}, {Castellano}, {Cleri}, {Cooper}, {Croton}, {Daddi}, {Dekel},
  {Elbaz}, {Franco}, {Gawiser}, {Holwerda}, {Huertas-Company}, {Jaskot},
  {Leung}, {Lucas}, {Mobasher}, {Pandya}, {Tacchella}, {Weiner}, \&
  {Zavala}}]{Finkelstein2022arXiv221105792F}
{Finkelstein}, S.~L., {Bagley}, M.~B., {Ferguson}, H.~C., {et~al.} 2023, \apjl,
  946, L13

\bibitem[{{Flesch}(2021)}]{Flesch2021}
{Flesch}, E.~W. 2021, arXiv e-prints, arXiv:2105.12985

\bibitem[{{Freese} {et~al.}(2016){Freese}, {Rindler-Daller}, {Spolyar}, \&
  {Valluri}}]{Freese2016RPPh...79f6902F}
{Freese}, K., {Rindler-Daller}, T., {Spolyar}, D., \& {Valluri}, M. 2016,
  Reports on Progress in Physics, 79, 066902

\bibitem[{{Fujimoto} {et~al.}(2022){Fujimoto}, {Brammer}, {Watson}, {Magdis},
  {Kokorev}, {Greve}, {Toft}, {Walter}, {Valiante}, {Ginolfi}, {Schneider},
  {Valentino}, {Colina}, {Vestergaard}, {Marques-Chaves}, {Fynbo}, {Krips},
  {Steinhardt}, {Cortzen}, {Rizzo}, \& {Oesch}}]{Fujimoto2022Natur.604..261F}
{Fujimoto}, S., {Brammer}, G.~B., {Watson}, D., {et~al.} 2022, \nat, 604, 261

\bibitem[{{Furtak} {et~al.}(2023{\natexlab{a}}){Furtak}, {Shuntov}, {Atek},
  {Zitrin}, {Richard}, {Lehnert}, \& {Chevallard}}]{furtak22b}
{Furtak}, L.~J., {Shuntov}, M., {Atek}, H., {et~al.} 2023{\natexlab{a}},
  \mnras, 519, 3064

\bibitem[{{Furtak} {et~al.}(2022){Furtak}, {Zitrin}, {Weaver}, {Atek},
  {Bezanson}, {Labbe}, {Whitaker}, {Leja}, {Price}, {Brammer}, {Wang}, {Dayal},
  {van Dokkum}, {Feldmann}, {Franx}, {Nelson}, \& {Mowla}}]{Furtak2022UNCOVER}
{Furtak}, L.~J., {Zitrin}, A., {Weaver}, J.~R., {et~al.} 2022, arXiv e-prints,
  arXiv:2212.04381

\bibitem[{{Furtak} {et~al.}(2023{\natexlab{b}}){Furtak}, {Mainali}, {Zitrin},
  {Plat}, {Fujimoto}, {Donahue}, {Nelson}, {Bauer}, {Uematsu}, {Caminha},
  {Andrade-Santos}, {Bradley}, {Caputi}, {Charlot}, {Chevallard}, {Coe},
  {Curtis-Lake}, {Espada}, {Frye}, {Knudsen}, {Koekemoer}, {Kohno}, {Kokorev},
  {Laporte}, {Lee}, {Lemaux}, {Magdis}, {Sharon}, {Stark}, {Su}, {Suess},
  {Ueda}, {Umehata}, {Vidal-Garc{\'\i}a}, \& {Wu}}]{Furtak2023AGN}
{Furtak}, L.~J., {Mainali}, R., {Zitrin}, A., {et~al.} 2023{\natexlab{b}},
  \mnras, 522, 5142

\bibitem[{{Gardner} {et~al.}(2006){Gardner}, {Mather}, {Clampin}, {Doyon},
  {Greenhouse}, {Hammel}, {Hutchings}, {Jakobsen}, {Lilly}, {Long}, {Lunine},
  {McCaughrean}, {Mountain}, {Nella}, {Rieke}, {Rieke}, {Rix}, {Smith},
  {Sonneborn}, {Stiavelli}, {Stockman}, {Windhorst}, \& {Wright}}]{Gardner2006}
{Gardner}, J.~P., {Mather}, J.~C., {Clampin}, M., {et~al.} 2006, \ssr, 123, 485

\bibitem[{{Gieles} {et~al.}(2018){Gieles}, {Charbonnel}, {Krause},
  {H{\'e}nault-Brunet}, {Agertz}, {Lamers}, {Bastian}, {Gualandris}, {Zocchi},
  \& {Petts}}]{Gieles2018MNRAS.478.2461G}
{Gieles}, M., {Charbonnel}, C., {Krause}, M. G.~H., {et~al.} 2018, \mnras, 478,
  2461

\bibitem[{{Glikman} {et~al.}(2011){Glikman}, {Djorgovski}, {Stern}, {Dey},
  {Jannuzi}, \& {Lee}}]{Glikman2011ApJ...728L..26G}
{Glikman}, E., {Djorgovski}, S.~G., {Stern}, D., {et~al.} 2011, \apjl, 728, L26

\bibitem[{{Glikman} {et~al.}(2006){Glikman}, {Helfand}, \&
  {White}}]{Glikman2006AGNtemplate}
{Glikman}, E., {Helfand}, D.~J., \& {White}, R.~L. 2006, \apj, 640, 579

\bibitem[{{Glikman} {et~al.}(2013){Glikman}, {Urrutia}, {Lacy}, {Djorgovski},
  {Urry}, {Croom}, {Schneider}, {Mahabal}, {Graham}, \& {Ge}}]{Glikman2013}
{Glikman}, E., {Urrutia}, T., {Lacy}, M., {et~al.} 2013, \apj, 778, 127

\bibitem[{{Glikman} {et~al.}(2023){Glikman}, {Rusu}, {Chen}, {Chan},
  {Spingola}, {Stacey}, {McKean}, {Berghea}, {Djorgovski}, {Graham}, {Stern},
  {Urrutia}, {Lacy}, {Secrest}, \& {O'Meara}}]{Glikman2023}
{Glikman}, E., {Rusu}, C.~E., {Chen}, G. C.~F., {et~al.} 2023, \apj, 943, 25

\bibitem[{{Gonz{\'a}lez-L{\'o}pez} {et~al.}(2017){Gonz{\'a}lez-L{\'o}pez},
  {Bauer}, {Romero-Ca{\~n}izales}, {Kneissl}, {Villard}, {Carvajal}, {Kim},
  {Laporte}, {Anguita}, {Aravena}, {Bouwens}, {Bradley}, {Carrasco}, {Demarco},
  {Ford}, {Ibar}, {Infante}, {Messias}, {Mu{\~n}oz Arancibia}, {Nagar},
  {Padilla}, {Treister}, {Troncoso}, \& {Zitrin}}]{gonzales-lopez17}
{Gonz{\'a}lez-L{\'o}pez}, J., {Bauer}, F.~E., {Romero-Ca{\~n}izales}, C.,
  {et~al.} 2017, \aap, 597, A41

\bibitem[{{Gutkin} {et~al.}(2016){Gutkin}, {Charlot}, \& {Bruzual}}]{gutkin16}
{Gutkin}, J., {Charlot}, S., \& {Bruzual}, G. 2016, \mnras, 462, 1757

\bibitem[{{Harikane} {et~al.}(2023){Harikane}, {Zhang}, {Nakajima}, {Ouchi},
  {Isobe}, {Ono}, {Hatano}, {Xu}, \& {Umeda}}]{Harikane2023}
{Harikane}, Y., {Zhang}, Y., {Nakajima}, K., {et~al.} 2023, arXiv e-prints,
  arXiv:2303.11946

\bibitem[{{Hsiao} {et~al.}(2022){Hsiao}, {Coe}, {Abdurro'uf}, {Whitler},
  {Jung}, {Khullar}, {Meena}, {Dayal}, {Barrow}, {Santos-Olmsted}, {Casselman},
  {Vanzella}, {Nonino}, {Jimenez-Teja}, {Oguri}, {Stark}, {Furtak}, {Zitrin},
  {Adamo}, {Brammer}, {Bradley}, {Diego}, {Zackrisson}, {Finkelstein},
  {Windhorst}, {Bhatawdekar}, {Hutchison}, {Broadhurst}, {Dimauro},
  {Andrade-Santos}, {Eldridge}, {Acebron}, {Avila}, {Bayliss}, {Benitez},
  {Binggeli}, {Bolan}, {Bradac}, {Carnall}, {Conselice}, {Donahue}, {Frye},
  {Fujimoto}, {Henry}, {James}, {Kassin}, {Kewley}, {Larson}, {Lauer}, {Law},
  {Mahler}, {Mainali}, {McCandliss}, {Nicholls}, {Pirzkal}, {Postman}, {Rigby},
  {Ryan}, {Senchyna}, {Sharon}, {Shimizu}, {Strait}, {Tang}, {Trenti},
  {Vikaeus}, \& {Welch}}]{Hsiao2022arXiv221014123H}
{Hsiao}, T. Y.-Y., {Coe}, D., {Abdurro'uf}, {et~al.} 2022, arXiv e-prints,
  arXiv:2210.14123

\bibitem[{Hunter(2007)}]{hunter07}
Hunter, J.~D. 2007, Computing in Science \& Engineering, 9, 90

\bibitem[{{Inada} {et~al.}(2003){Inada}, {Oguri}, {Pindor}, {Hennawi}, {Chiu},
  {Zheng}, {Ichikawa}, {Gregg}, {Becker}, {Suto}, {Strauss}, {Turner},
  {Keeton}, {Annis}, {Castander}, {Eisenstein}, {Frieman}, {Fukugita}, {Gunn},
  {Johnston}, {Kent}, {Nichol}, {Richards}, {Rix}, {Sheldon}, {Bahcall},
  {Brinkmann}, {Ivezi{\'c}}, {Lamb}, {McKay}, {Schneider}, \&
  {York}}]{Inada2003Natur.426..810I}
{Inada}, N., {Oguri}, M., {Pindor}, B., {et~al.} 2003, \nat, 426, 810

\bibitem[{{Inoue} {et~al.}(2014){Inoue}, {Shimizu}, {Iwata}, \&
  {Tanaka}}]{inoue14}
{Inoue}, A.~K., {Shimizu}, I., {Iwata}, I., \& {Tanaka}, M. 2014, \mnras, 442,
  1805

\bibitem[{{Jakobsen} {et~al.}(2022){Jakobsen}, {Ferruit}, {Alves de Oliveira},
  {Arribas}, {Bagnasco}, {Barho}, {Beck}, {Birkmann}, {B{\"o}ker}, {Bunker},
  {Charlot}, {de Jong}, {de Marchi}, {Ehrenwinkler}, {Falcolini}, {Fels},
  {Franx}, {Franz}, {Funke}, {Giardino}, {Gnata}, {Holota}, {Honnen}, {Jensen},
  {Jentsch}, {Johnson}, {Jollet}, {Karl}, {Kling}, {K{\"o}hler}, {Kolm},
  {Kumari}, {Lander}, {Lemke}, {L{\'o}pez-Caniego}, {L{\"u}tzgendorf},
  {Maiolino}, {Manjavacas}, {Marston}, {Maschmann}, {Maurer}, {Messerschmidt},
  {Moseley}, {Mosner}, {Mott}, {Muzerolle}, {Pirzkal}, {Pittet}, {Plitzke},
  {Posselt}, {Rapp}, {Rauscher}, {Rawle}, {Rix}, {R{\"o}del}, {Rumler},
  {Sabbi}, {Salvignol}, {Schmid}, {Sirianni}, {Smith}, {Strada}, {te Plate},
  {Valenti}, {Wettemann}, {Wiehe}, {Wiesmayer}, {Willott}, {Wright}, {Zeidler},
  \& {Zincke}}]{jakobsen22}
{Jakobsen}, P., {Ferruit}, P., {Alves de Oliveira}, C., {et~al.} 2022, \aap,
  661, A80

\bibitem[{{Johnson} {et~al.}(2021){Johnson}, {Leja}, {Conroy}, \&
  {Speagle}}]{prospector21}
{Johnson}, B.~D., {Leja}, J., {Conroy}, C., \& {Speagle}, J.~S. 2021, \apjs,
  254, 22

\bibitem[{{Kawamata} {et~al.}(2018){Kawamata}, {Ishigaki}, {Shimasaku},
  {Oguri}, {Ouchi}, \& {Tanigawa}}]{Kawamata2018ApJ...855....4K}
{Kawamata}, R., {Ishigaki}, M., {Shimasaku}, K., {et~al.} 2018, \apj, 855, 4

\bibitem[{{Keeton}(2001)}]{Keeton2001models}
{Keeton}, C.~R. 2001, ArXiv Astrophysics e-prints, arXiv:astro-ph/0102340

\bibitem[{{Kocevski} {et~al.}(2023){Kocevski}, {Onoue}, {Inayoshi}, {Trump},
  {Arrabal Haro}, {Grazian}, {Dickinson}, {Finkelstein}, {Kartaltepe},
  {Hirschmann}, {Fujimoto}, {Juneau}, {Amorin}, {Bagley}, {Barro}, {Bell},
  {Bisigello}, {Calabro}, {Cleri}, {Cooper}, {Ding}, {Grogin}, {Ho}, {Inoue},
  {Jiang}, {Jones}, {Koekemoer}, {Li}, {Li}, {McGrath}, {Molina}, {Papovich},
  {Perez-Gonzalez}, {Pirzkal}, {Wilkins}, {Yang}, \& {Yung}}]{Kocevski2023}
{Kocevski}, D.~D., {Onoue}, M., {Inayoshi}, K., {et~al.} 2023, arXiv e-prints,
  arXiv:2302.00012

\bibitem[{{Kohno}(2019)}]{Kohno2019asrc.confE..64K}
{Kohno}, K. 2019, in ALMA2019: Science Results and Cross-Facility Synergies, 64

\bibitem[{{Labb{\'e}} {et~al.}(2013){Labb{\'e}}, {Oesch}, {Bouwens},
  {Illingworth}, {Magee}, {Gonz{\'a}lez}, {Carollo}, {Franx}, {Trenti}, {van
  Dokkum}, \& {Stiavelli}}]{Labbe2013}
{Labb{\'e}}, I., {Oesch}, P.~A., {Bouwens}, R.~J., {et~al.} 2013, \apjl, 777,
  L19

\bibitem[{{Labb{\'e}} {et~al.}(2023){Labb{\'e}}, {van Dokkum}, {Nelson},
  {Bezanson}, {Suess}, {Leja}, {Brammer}, {Whitaker}, {Mathews}, {Stefanon}, \&
  {Wang}}]{Labbe2022arXiv220712446L}
{Labb{\'e}}, I., {van Dokkum}, P., {Nelson}, E., {et~al.} 2023, \nat, 616, 266

\bibitem[{{Lam} {et~al.}(2014){Lam}, {Broadhurst}, {Diego}, {Lim}, {Coe},
  {Ford}, \& {Zheng}}]{Lam2014modelA2744}
{Lam}, D., {Broadhurst}, T., {Diego}, J.~M., {et~al.} 2014, \apj, 797, 98

\bibitem[{{Laporte} {et~al.}(2017){Laporte}, {Ellis}, {Boone}, {Bauer},
  {Qu{\'e}nard}, {Roberts-Borsani}, {Pell{\'o}}, {P{\'e}rez-Fournon}, \&
  {Streblyanska}}]{Laporte2017}
{Laporte}, N., {Ellis}, R.~S., {Boone}, F., {et~al.} 2017, \apjl, 837, L21

\bibitem[{{Larson} {et~al.}(2023){Larson}, {Finkelstein}, {Kocevski},
  {Hutchison}, {Trump}, {Arrabal Haro}, {Bromm}, {Cleri}, {Dickinson},
  {Fujimoto}, {Kartaltepe}, {Koekemoer}, {Papovich}, {Pirzkal}, {Tacchella},
  {Zavala}, {Bagley}, {Behroozi}, {Champagne}, {Cole}, {Jung}, {Morales},
  {Yang}, {Zhang}, {Zitrin}, {Amor{\'\i}n}, {Burgarella}, {Casey}, {Ch{\'a}vez
  Ortiz}, {Cox}, {Chworowsky}, {Fontana}, {Gawiser}, {Grazian}, {Grogin},
  {Harish}, {Hathi}, {Hirschmann}, {Holwerda}, {Juneau}, {Leung}, {Lucas},
  {McGrath}, {P{\'e}rez-Gonz{\'a}lez}, {Rigby}, {Seill{\'e}}, {Simons},
  {Weiner}, {Wilkins}, {Yung}, \& {The CEERS Team}}]{Larson2023}
{Larson}, R.~L., {Finkelstein}, S.~L., {Kocevski}, D.~D., {et~al.} 2023, arXiv
  e-prints, arXiv:2303.08918

\bibitem[{{Leitherer} {et~al.}(2002){Leitherer}, {Li}, {Calzetti}, \&
  {Heckman}}]{Leitherer2002}
{Leitherer}, C., {Li}, I.~H., {Calzetti}, D., \& {Heckman}, T.~M. 2002, \apjs,
  140, 303

\bibitem[{{Leja} {et~al.}(2019){Leja}, {Carnall}, {Johnson}, {Conroy}, \&
  {Speagle}}]{Leja2019}
{Leja}, J., {Carnall}, A.~C., {Johnson}, B.~D., {Conroy}, C., \& {Speagle},
  J.~S. 2019, \apj, 876, 3

\bibitem[{{Leja} {et~al.}(2018){Leja}, {Johnson}, {Conroy}, \& {van
  Dokkum}}]{Leja2018ProspectorAGN}
{Leja}, J., {Johnson}, B.~D., {Conroy}, C., \& {van Dokkum}, P. 2018, \apj,
  854, 62

\bibitem[{{Leja} {et~al.}(2017){Leja}, {Johnson}, {Conroy}, {van Dokkum}, \&
  {Byler}}]{Leja2017}
{Leja}, J., {Johnson}, B.~D., {Conroy}, C., {van Dokkum}, P.~G., \& {Byler}, N.
  2017, \apj, 837, 170

\bibitem[{{Leja} {et~al.}(2020){Leja}, {Speagle}, {Johnson}, {Conroy}, {van
  Dokkum}, \& {Franx}}]{Leja2020}
{Leja}, J., {Speagle}, J.~S., {Johnson}, B.~D., {et~al.} 2020, \apj, 893, 111

\bibitem[{{Lotz} {et~al.}(2017){Lotz}, {Koekemoer}, {Coe}, {Grogin}, {Capak},
  {Mack}, {Anderson}, {Avila}, {Barker}, {Borncamp}, {Brammer}, {Durbin},
  {Gunning}, {Hilbert}, {Jenkner}, {Khandrika}, {Levay}, {Lucas}, {MacKenty},
  {Ogaz}, {Porterfield}, {Reid}, {Robberto}, {Royle}, {Smith},
  {Storrie-Lombardi}, {Sunnquist}, {Surace}, {Taylor}, {Williams}, {Bullock},
  {Dickinson}, {Finkelstein}, {Natarajan}, {Richard}, {Robertson}, {Tumlinson},
  {Zitrin}, {Flanagan}, {Sembach}, {Soifer}, \& {Mountain}}]{Lotz2016HFF}
{Lotz}, J.~M., {Koekemoer}, A., {Coe}, D., {et~al.} 2017, \apj, 837, 97

\bibitem[{{Lyke} {et~al.}(2020){Lyke}, {Higley}, {McLane}, {Schurhammer},
  {Myers}, {Ross}, {Dawson}, {Chabanier}, {Martini}, {Busca}, {Mas des
  Bourboux}, {Salvato}, {Streblyanska}, {Zarrouk}, {Burtin}, {Anderson},
  {Bautista}, {Bizyaev}, {Brandt}, {Brinkmann}, {Brownstein}, {Comparat},
  {Green}, {de la Macorra}, {Mu{\~n}oz Guti{\'e}rrez}, {Hou}, {Newman},
  {Palanque-Delabrouille}, {P{\^a}ris}, {Percival}, {Petitjean}, {Rich},
  {Rossi}, {Schneider}, {Smith}, {Vivek}, \& {Weaver}}]{Lyke2020}
{Lyke}, B.~W., {Higley}, A.~N., {McLane}, J.~N., {et~al.} 2020, \apjs, 250, 8

\bibitem[{{Mahler} {et~al.}(2018){Mahler}, {Richard}, {Cl{\'e}ment},
  {Lagattuta}, {Schmidt}, {Patr{\'\i}cio}, {Soucail}, {Bacon}, {Pello},
  {Bouwens}, {Maseda}, {Martinez}, {Carollo}, {Inami}, {Leclercq}, \&
  {Wisotzki}}]{Mahler2018A2744}
{Mahler}, G., {Richard}, J., {Cl{\'e}ment}, B., {et~al.} 2018, \mnras, 473, 663

\bibitem[{{Mainali} {et~al.}(2017){Mainali}, {Kollmeier}, {Stark}, {Simcoe},
  {Walth}, {Newman}, \& {Miller}}]{Mainali2017}
{Mainali}, R., {Kollmeier}, J.~A., {Stark}, D.~P., {et~al.} 2017, \apjl, 836,
  L14

\bibitem[{{Martins} {et~al.}(2020){Martins}, {Schaerer}, {Haemmerl{\'e}}, \&
  {Charbonnel}}]{Martins2020SupermassiveStars}
{Martins}, F., {Schaerer}, D., {Haemmerl{\'e}}, L., \& {Charbonnel}, C. 2020,
  \aap, 633, A9

\bibitem[{{Matthee}(2021)}]{Matthee2021colors}
{Matthee}, J. 2021, Nature Astronomy, 5, 984

\bibitem[{{Matthee} {et~al.}(2020){Matthee}, {Pezzulli}, {Mackenzie},
  {Cantalupo}, {Kusakabe}, {Leclercq}, {Sobral}, {Richard}, {Wisotzki},
  {Lilly}, {Boogaard}, {Marino}, {Maseda}, \& {Nanayakkara}}]{Matthee2020CR7}
{Matthee}, J., {Pezzulli}, G., {Mackenzie}, R., {et~al.} 2020, \mnras, 498,
  3043

\bibitem[{{McCracken} {et~al.}(2012){McCracken}, {Milvang-Jensen}, {Dunlop},
  {Franx}, {Fynbo}, {Le F{\`e}vre}, {Holt}, {Caputi}, {Goranova}, {Buitrago},
  {Emerson}, {Freudling}, {Hudelot}, {L{\'o}pez-Sanjuan}, {Magnard}, {Mellier},
  {M{\o}ller}, {Nilsson}, {Sutherland}, {Tasca}, \&
  {Zabl}}]{Ultravista2012A&A...544A.156M}
{McCracken}, H.~J., {Milvang-Jensen}, B., {Dunlop}, J., {et~al.} 2012, \aap,
  544, A156

\bibitem[{{McElwain} {et~al.}(2023){McElwain}, {Feinberg}, {Perrin}, {Clampin},
  {Mountain}, {Lallo}, {Lajoie}, {Kimble}, {Bowers}, {Stark}, {Acton},
  {Atkinson}, {Barinek}, {Barto}, {Basinger}, {Beck}, {Bergkoetter}, {Bluth},
  {Boucarut}, {Brady}, {Brooks}, {Brown}, {Byard}, {Carey}, {Carrasquilla},
  {Chae}, {Chaney}, {Chayer}, {Chonis}, {Cohen}, {Cole}, {Comeau}, {Coon},
  {Coppock}, {Coyle}, {Dean}, {Dziak}, {Eisenhower}, {Flagey}, {Franck},
  {Gallagher}, {Gilman}, {Glassman}, {Green}, {Grieco}, {Haase},
  {Hadjimichael}, {Hagopian}, {Hahn}, {Hartig}, {Havey}, {Hayden}, {Hellekson},
  {Hicks}, {Holfeltz}, {Howard}, {Huguet}, {Jahne}, {Johnson}, {Johnston},
  {Jurling}, {Kegley}, {Kennard}, {Keski-Kuha}, {Knight}, {Kulp}, {Levi},
  {Levine}, {Lightsey}, {Luetgens}, {Mather}, {Matthews}, {McKay}, {Mehalick},
  {Mel{\'e}ndez}, {Mosier}, {Murphy}, {Nelan}, {Niedner}, {Nol}, {Ohara},
  {Ohl}, {Olczak}, {Osborne}, {Park}, {Perrygo}, {Pueyo}, {Redding}, {Regan},
  {Reynolds}, {Rifelli}, {Rigby}, {Sabatke}, {Saif}, {Scorse}, {Seo}, {Shi},
  {Sigrist}, {Smith}, {Smith}, {Smith}, {Sohn}, {Stahl}, {Telfer}, {Terlecki},
  {Texter}, {Van Buren}, {Van Campen}, {Vila}, {Voyton}, {Waldman}, {Walker},
  {Weiser}, {Wells}, {West}, {Whitman}, {Wolf}, \& {Zielinski}}]{McElwain2023}
{McElwain}, M.~W., {Feinberg}, L.~D., {Perrin}, M.~D., {et~al.} 2023, \pasp,
  135, 058001

\bibitem[{{Meiksin}(2006)}]{Meiksin2006}
{Meiksin}, A. 2006, \mnras, 365, 807

\bibitem[{{Merten} {et~al.}(2011){Merten}, {Coe}, {Dupke}, {Massey}, {Zitrin},
  {Cypriano}, {Okabe}, {Frye}, {Braglia}, {Jim{\'e}nez-Teja}, {Ben{\'{\i}}tez},
  {Broadhurst}, {Rhodes}, {Meneghetti}, {Moustakas}, {Sodr{\'e}}, {Krick}, \&
  {Bregman}}]{Merten2011}
{Merten}, J., {Coe}, D., {Dupke}, R., {et~al.} 2011, \mnras, 417, 333

\bibitem[{{Morishita} {et~al.}(2023){Morishita}, {Roberts-Borsani}, {Treu},
  {Brammer}, {Mason}, {Trenti}, {Vulcani}, {Wang}, {Acebron}, {Bah{\'e}},
  {Bergamini}, {Boyett}, {Bradac}, {Calabr{\`o}}, {Castellano}, {Chen}, {De
  Lucia}, {Filippenko}, {Fontana}, {Glazebrook}, {Grillo}, {Henry}, {Jones},
  {Kelly}, {Koekemoer}, {Leethochawalit}, {Lu}, {Marchesini}, {Mascia},
  {Mercurio}, {Merlin}, {Metha}, {Nanayakkara}, {Nonino}, {Paris},
  {Pentericci}, {Rosati}, {Santini}, {Strait}, {Vanzella}, {Windhorst}, \&
  {Xie}}]{Morishita2022arXiv221109097M}
{Morishita}, T., {Roberts-Borsani}, G., {Treu}, T., {et~al.} 2023, \apjl, 947,
  L24

\bibitem[{{Murphy} {et~al.}(2017){Murphy}, {Momjian}, {Condon}, {Chary},
  {Dickinson}, {Inami}, {Taylor}, \& {Weiner}}]{Murphy2017ApJ...839...35M}
{Murphy}, E.~J., {Momjian}, E., {Condon}, J.~J., {et~al.} 2017, \apj, 839, 35

\bibitem[{Niida {et~al.}(2020)Niida, Nagao, Ikeda, Akiyama, Matsuoka, He,
  Matsuoka, Toba, Onoue, Kobayashi, Taniguchi, Furusawa, Harikane, Imanishi,
  Kashikawa, Kawaguchi, Komiyama, Shirakata, Terashima, \& Ueda}]{Niida_2020}
Niida, M., Nagao, T., Ikeda, H., {et~al.} 2020, The Astrophysical Journal, 904,
  89.
\newblock \url{https://dx.doi.org/10.3847/1538-4357/abbe11}

\bibitem[{{Oesch} {et~al.}(2015){Oesch}, {van Dokkum}, {Illingworth},
  {Bouwens}, {Momcheva}, {Holden}, {Roberts-Borsani}, {Smit}, {Franx},
  {Labb{\'e}}, {Gonz{\'a}lez}, \& {Magee}}]{Oesch20157p73}
{Oesch}, P.~A., {van Dokkum}, P.~G., {Illingworth}, G.~D., {et~al.} 2015,
  \apjl, 804, L30

\bibitem[{{Ofek}(2014)}]{maat14}
{Ofek}, E.~O. 2014, {MAAT: MATLAB Astronomy and Astrophysics Toolbox},
  Astrophysics Source Code Library, record ascl:1407.005, , , ascl:1407.005

\bibitem[{{Oguri} {et~al.}(2013){Oguri}, {Schrabback}, {Jullo}, {Ota},
  {Kochanek}, {Dai}, {Ofek}, {Richards}, {Blandford}, {Falco}, \&
  {Fohlmeister}}]{Oguri2013MNRAS.429..482O}
{Oguri}, M., {Schrabback}, T., {Jullo}, E., {et~al.} 2013, \mnras, 429, 482

\bibitem[{{Oke} \& {Gunn}(1983)}]{Oke1983ABandStandards}
{Oke}, J.~B., \& {Gunn}, J.~E. 1983, \apj, 266, 713

\bibitem[{{Onoue} {et~al.}(2023){Onoue}, {Inayoshi}, {Ding}, {Li}, {Li},
  {Molina}, {Inoue}, {Jiang}, \& {Ho}}]{Onoue2022arXiv220907325O}
{Onoue}, M., {Inayoshi}, K., {Ding}, X., {et~al.} 2023, \apjl, 942, L17

\bibitem[{{Pacucci} {et~al.}(2016){Pacucci}, {Ferrara}, {Grazian}, {Fiore},
  {Giallongo}, \& {Puccetti}}]{Pacucci2016DCBH}
{Pacucci}, F., {Ferrara}, A., {Grazian}, A., {et~al.} 2016, \mnras, 459, 1432

\bibitem[{{Padmanabhan} \& {Loeb}(2022)}]{PadmanabhanLoeb2022GReGr..54...24P}
{Padmanabhan}, H., \& {Loeb}, A. 2022, General Relativity and Gravitation, 54,
  24

\bibitem[{{Pei}(1992)}]{pei92}
{Pei}, Y.~C. 1992, \apj, 395, 130

\bibitem[{{Peng} {et~al.}(2010){Peng}, {Ho}, {Impey}, \&
  {Rix}}]{Peng2010AJ....139.2097P}
{Peng}, C.~Y., {Ho}, L.~C., {Impey}, C.~D., \& {Rix}, H.-W. 2010, \aj, 139,
  2097

\bibitem[{{Pentericci} {et~al.}(2011){Pentericci}, {Fontana}, {Vanzella},
  {Castellano}, {Grazian}, {Dijkstra}, {Boutsia}, {Cristiani}, {Dickinson},
  {Giallongo}, {Giavalisco}, {Maiolino}, {Moorwood}, {Paris}, \&
  {Santini}}]{Pentericci2011Lyfractionz7}
{Pentericci}, L., {Fontana}, A., {Vanzella}, E., {et~al.} 2011, \apj, 743, 132

\bibitem[{{Planck Collaboration} {et~al.}(2016){Planck Collaboration}, {Ade},
  {Aghanim}, {Arnaud}, {Ashdown}, {Aumont}, {Baccigalupi}, {Banday},
  {Barreiro}, {Bartlett}, {Bartolo}, {Battaner}, {Battye}, {Benabed},
  {Beno{\^\i}t}, {Benoit-L{\'e}vy}, {Bernard}, {Bersanelli}, {Bielewicz},
  {Bock}, {Bonaldi}, {Bonavera}, {Bond}, {Borrill}, {Bouchet}, {Boulanger},
  {Bucher}, {Burigana}, {Butler}, {Calabrese}, {Cardoso}, {Catalano},
  {Challinor}, {Chamballu}, {Chary}, {Chiang}, {Chluba}, {Christensen},
  {Church}, {Clements}, {Colombi}, {Colombo}, {Combet}, {Coulais}, {Crill},
  {Curto}, {Cuttaia}, {Danese}, {Davies}, {Davis}, {de Bernardis}, {de Rosa},
  {de Zotti}, {Delabrouille}, {D{\'e}sert}, {Di Valentino}, {Dickinson},
  {Diego}, {Dolag}, {Dole}, {Donzelli}, {Dor{\'e}}, {Douspis}, {Ducout},
  {Dunkley}, {Dupac}, {Efstathiou}, {Elsner}, {En{\ss}lin}, {Eriksen},
  {Farhang}, {Fergusson}, {Finelli}, {Forni}, {Frailis}, {Fraisse},
  {Franceschi}, {Frejsel}, {Galeotta}, {Galli}, {Ganga}, {Gauthier}, {Gerbino},
  {Ghosh}, {Giard}, {Giraud-H{\'e}raud}, {Giusarma}, {Gjerl{\o}w},
  {Gonz{\'a}lez-Nuevo}, {G{\'o}rski}, {Gratton}, {Gregorio}, {Gruppuso},
  {Gudmundsson}, {Hamann}, {Hansen}, {Hanson}, {Harrison}, {Helou},
  {Henrot-Versill{\'e}}, {Hern{\'a}ndez-Monteagudo}, {Herranz}, {Hildebrandt},
  {Hivon}, {Hobson}, {Holmes}, {Hornstrup}, {Hovest}, {Huang}, {Huffenberger},
  {Hurier}, {Jaffe}, {Jaffe}, {Jones}, {Juvela}, {Keih{\"a}nen}, {Keskitalo},
  {Kisner}, {Kneissl}, {Knoche}, {Knox}, {Kunz}, {Kurki-Suonio}, {Lagache},
  {L{\"a}hteenm{\"a}ki}, {Lamarre}, {Lasenby}, {Lattanzi}, {Lawrence}, {Leahy},
  {Leonardi}, {Lesgourgues}, {Levrier}, {Lewis}, {Liguori}, {Lilje},
  {Linden-V{\o}rnle}, {L{\'o}pez-Caniego}, {Lubin}, {Mac{\'\i}as-P{\'e}rez},
  {Maggio}, {Maino}, {Mandolesi}, {Mangilli}, {Marchini}, {Maris}, {Martin},
  {Martinelli}, {Mart{\'\i}nez-Gonz{\'a}lez}, {Masi}, {Matarrese}, {McGehee},
  {Meinhold}, {Melchiorri}, {Melin}, {Mendes}, {Mennella}, {Migliaccio},
  {Millea}, {Mitra}, {Miville-Desch{\^e}nes}, {Moneti}, {Montier}, {Morgante},
  {Mortlock}, {Moss}, {Munshi}, {Murphy}, {Naselsky}, {Nati}, {Natoli},
  {Netterfield}, {N{\o}rgaard-Nielsen}, {Noviello}, {Novikov}, {Novikov},
  {Oxborrow}, {Paci}, {Pagano}, {Pajot}, {Paladini}, {Paoletti}, {Partridge},
  {Pasian}, {Patanchon}, {Pearson}, {Perdereau}, {Perotto}, {Perrotta},
  {Pettorino}, {Piacentini}, {Piat}, {Pierpaoli}, {Pietrobon}, {Plaszczynski},
  {Pointecouteau}, {Polenta}, {Popa}, {Pratt}, {Pr{\'e}zeau}, {Prunet},
  {Puget}, {Rachen}, {Reach}, {Rebolo}, {Reinecke}, {Remazeilles}, {Renault},
  {Renzi}, {Ristorcelli}, {Rocha}, {Rosset}, {Rossetti}, {Roudier},
  {Rouill{\'e} d'Orfeuil}, {Rowan-Robinson}, {Rubi{\~n}o-Mart{\'\i}n},
  {Rusholme}, {Said}, {Salvatelli}, {Salvati}, {Sandri}, {Santos},
  {Savelainen}, {Savini}, {Scott}, {Seiffert}, {Serra}, {Shellard}, {Spencer},
  {Spinelli}, {Stolyarov}, {Stompor}, {Sudiwala}, {Sunyaev}, {Sutton},
  {Suur-Uski}, {Sygnet}, {Tauber}, {Terenzi}, {Toffolatti}, {Tomasi},
  {Tristram}, {Trombetti}, {Tucci}, {Tuovinen}, {T{\"u}rler}, {Umana},
  {Valenziano}, {Valiviita}, {Van Tent}, {Vielva}, {Villa}, {Wade}, {Wandelt},
  {Wehus}, {White}, {White}, {Wilkinson}, {Yvon}, {Zacchei}, \&
  {Zonca}}]{Planck15}
{Planck Collaboration}, {Ade}, P.~A.~R., {Aghanim}, N., {et~al.} 2016, \aap,
  594, A13

\bibitem[{{Price-Whelan} {et~al.}(2018){Price-Whelan}, {Sip{\H{o}}cz},
  {G{\"u}nther}, {Lim}, {Crawford}, {Conseil}, {Shupe}, {Craig}, {Dencheva},
  {Ginsburg}, {VanderPlas}, {Bradley}, {P{\'e}rez-Su{\'a}rez}, {de Val-Borro},
  {Paper Contributors}, {Aldcroft}, {Cruz}, {Robitaille}, {Tollerud},
  {Coordination Committee}, {Ardelean}, {Babej}, {Bach}, {Bachetti}, {Bakanov},
  {Bamford}, {Barentsen}, {Barmby}, {Baumbach}, {Berry}, {Biscani}, {Boquien},
  {Bostroem}, {Bouma}, {Brammer}, {Bray}, {Breytenbach}, {Buddelmeijer},
  {Burke}, {Calderone}, {Cano Rodr{\'\i}guez}, {Cara}, {Cardoso}, {Cheedella},
  {Copin}, {Corrales}, {Crichton}, {D{\textquoteright}Avella}, {Deil},
  {Depagne}, {Dietrich}, {Donath}, {Droettboom}, {Earl}, {Erben}, {Fabbro},
  {Ferreira}, {Finethy}, {Fox}, {Garrison}, {Gibbons}, {Goldstein}, {Gommers},
  {Greco}, {Greenfield}, {Groener}, {Grollier}, {Hagen}, {Hirst}, {Homeier},
  {Horton}, {Hosseinzadeh}, {Hu}, {Hunkeler}, {Ivezi{\'c}}, {Jain}, {Jenness},
  {Kanarek}, {Kendrew}, {Kern}, {Kerzendorf}, {Khvalko}, {King}, {Kirkby},
  {Kulkarni}, {Kumar}, {Lee}, {Lenz}, {Littlefair}, {Ma}, {Macleod},
  {Mastropietro}, {McCully}, {Montagnac}, {Morris}, {Mueller}, {Mumford},
  {Muna}, {Murphy}, {Nelson}, {Nguyen}, {Ninan}, {N{\"o}the}, {Ogaz}, {Oh},
  {Parejko}, {Parley}, {Pascual}, {Patil}, {Patil}, {Plunkett}, {Prochaska},
  {Rastogi}, {Reddy Janga}, {Sabater}, {Sakurikar}, {Seifert}, {Sherbert},
  {Sherwood-Taylor}, {Shih}, {Sick}, {Silbiger}, {Singanamalla}, {Singer},
  {Sladen}, {Sooley}, {Sornarajah}, {Streicher}, {Teuben}, {Thomas},
  {Tremblay}, {Turner}, {Terr{\'o}n}, {van Kerkwijk}, {de la Vega}, {Watkins},
  {Weaver}, {Whitmore}, {Woillez}, {Zabalza}, \& {Contributors}}]{astropy18}
{Price-Whelan}, A.~M., {Sip{\H{o}}cz}, B.~M., {G{\"u}nther}, H.~M., {et~al.}
  2018, \aj, 156, 123

\bibitem[{{Rieke} {et~al.}(2005){Rieke}, {Kelly}, \& {Horner}}]{rieke05}
{Rieke}, M.~J., {Kelly}, D., \& {Horner}, S. 2005, Society of Photo-Optical
  Instrumentation Engineers (SPIE) Conference Series, Vol. 5904, {Overview of
  James Webb Space Telescope and NIRCam's Role}, ed. J.~B. {Heaney} \& L.~G.
  {Burriesci} (SPIE), 1--8

\bibitem[{{Rieke} {et~al.}(2023){Rieke}, {Kelly}, {Misselt}, {Stansberry},
  {Boyer}, {Beatty}, {Egami}, {Florian}, {Greene}, {Hainline}, {Leisenring},
  {Roellig}, {Schlawin}, {Sun}, {Tinnin}, {Williams}, {Willmer}, {Wilson},
  {Clark}, {Rohrbach}, {Brooks}, {Canipe}, {Correnti}, {DiFelice}, {Gennaro},
  {Girard}, {Hartig}, {Hilbert}, {Koekemoer}, {Nikolov}, {Pirzkal}, {Rest},
  {Robberto}, {Sunnquist}, {Telfer}, {Wu}, {Ferry}, {Lewis}, {Baum},
  {Beichman}, {Doyon}, {Dressler}, {Eisenstein}, {Ferrarese}, {Hodapp},
  {Horner}, {Jaffe}, {Johnstone}, {Krist}, {Martin}, {McCarthy}, {Meyer},
  {Rieke}, {Trauger}, \& {Young}}]{Rieke2023}
{Rieke}, M.~J., {Kelly}, D.~M., {Misselt}, K., {et~al.} 2023, \pasp, 135,
  028001

\bibitem[{{Roberts-Borsani} {et~al.}(2022){Roberts-Borsani}, {Treu}, {Chen},
  {Morishita}, {Vanzella}, {Zitrin}, {Bergamini}, {Castellano}, {Fontana},
  {Grillo}, {Kelly}, {Merlin}, {Paris}, {Rosati}, {Acebron}, {Bonchi},
  {Boyett}, {Bradac}, {Broadhurst}, {Calabro}, {Diego}, {Dressler}, {Furtak},
  {Filippenko}, {Glazebrook}, {Koekemoer}, {Leethochawalit}, {Malkan}, {Mason},
  {Mercurio}, {Metha}, {Nanayakkara}, {Pentericci}, {Pierel}, {Rieck}, {Roy},
  {Santini}, {Strait}, {Strausbaugh}, {Trenti}, {Vulcani}, {Wang}, {Wang},
  {Windhorst}, \& {Yang}}]{Roberts-Borsani2022arXiv221015639R}
{Roberts-Borsani}, G., {Treu}, T., {Chen}, W., {et~al.} 2022, arXiv e-prints,
  arXiv:2210.15639

\bibitem[{{Roberts-Borsani} {et~al.}(2016){Roberts-Borsani}, {Bouwens},
  {Oesch}, {Labbe}, {Smit}, {Illingworth}, {van Dokkum}, {Holden}, {Gonzalez},
  {Stefanon}, {Holwerda}, \& {Wilkins}}]{Roberts_Borsani2015}
{Roberts-Borsani}, G.~W., {Bouwens}, R.~J., {Oesch}, P.~A., {et~al.} 2016,
  \apj, 823, 143

\bibitem[{{Robertson} {et~al.}(2015){Robertson}, {Ellis}, {Furlanetto}, \&
  {Dunlop}}]{Robertson15}
{Robertson}, B.~E., {Ellis}, R.~S., {Furlanetto}, S.~R., \& {Dunlop}, J.~S.
  2015, \apjl, 802, L19

\bibitem[{{Ryon} {et~al.}(2017){Ryon}, {Gallagher}, {Smith}, {Adamo},
  {Calzetti}, {Bright}, {Cignoni}, {Cook}, {Dale}, {Elmegreen}, {Fumagalli},
  {Gouliermis}, {Grasha}, {Grebel}, {Kim}, {Messa}, {Thilker}, \&
  {Ubeda}}]{Ryon2017ApJ...841...92R}
{Ryon}, J.~E., {Gallagher}, J.~S., {Smith}, L.~J., {et~al.} 2017, \apj, 841, 92

\bibitem[{{Selsing} {et~al.}(2016){Selsing}, {Fynbo}, {Christensen}, \&
  {Krogager}}]{Selsing2016}
{Selsing}, J., {Fynbo}, J.~P.~U., {Christensen}, L., \& {Krogager}, J.~K. 2016,
  \aap, 585, A87

\bibitem[{{Sharon} {et~al.}(2017){Sharon}, {Bayliss}, {Dahle}, {Florian},
  {Gladders}, {Johnson}, {Paterno-Mahler}, {Rigby}, {Whitaker}, \&
  {Wuyts}}]{sharon2017}
{Sharon}, K., {Bayliss}, M.~B., {Dahle}, H., {et~al.} 2017, \apj, 835, 5

\bibitem[{{Shen} {et~al.}(2020){Shen}, {Hopkins}, {Faucher-Gigu{\`e}re},
  {Alexander}, {Richards}, {Ross}, \& {Hickox}}]{Shen2020QuasarLF}
{Shen}, X., {Hopkins}, P.~F., {Faucher-Gigu{\`e}re}, C.-A., {et~al.} 2020,
  \mnras, 495, 3252

\bibitem[{{Smidt} {et~al.}(2014){Smidt}, {Whalen}, {Wiggins}, {Even},
  {Johnson}, \& {Fryer}}]{Smidt2014ApJ...797...97S}
{Smidt}, J., {Whalen}, D.~J., {Wiggins}, B.~K., {et~al.} 2014, \apj, 797, 97

\bibitem[{{Smit} {et~al.}(2015){Smit}, {Bouwens}, {Franx}, {Oesch}, {Ashby},
  {Willner}, {Labb{\'e}}, {Holwerda}, {Fazio}, \&
  {Huang}}]{Smit2015ApJ...801..122S}
{Smit}, R., {Bouwens}, R.~J., {Franx}, M., {et~al.} 2015, \apj, 801, 122

\bibitem[{{Stalevski} {et~al.}(2012){Stalevski}, {Fritz}, {Baes}, {Nakos}, \&
  {Popovi{\'c}}}]{Stalevski2012}
{Stalevski}, M., {Fritz}, J., {Baes}, M., {Nakos}, T., \& {Popovi{\'c}},
  L.~{\v{C}}. 2012, \mnras, 420, 2756

\bibitem[{{Stalevski} {et~al.}(2016){Stalevski}, {Ricci}, {Ueda}, {Lira},
  {Fritz}, \& {Baes}}]{Stalevski2016}
{Stalevski}, M., {Ricci}, C., {Ueda}, Y., {et~al.} 2016, \mnras, 458, 2288

\bibitem[{{Stark} {et~al.}(2010){Stark}, {Ellis}, {Chiu}, {Ouchi}, \&
  {Bunker}}]{Stark2010z3-7fractions}
{Stark}, D.~P., {Ellis}, R.~S., {Chiu}, K., {Ouchi}, M., \& {Bunker}, A. 2010,
  \mnras, 408, 1628

\bibitem[{{Stark} {et~al.}(2015{\natexlab{a}}){Stark}, {Richard}, {Charlot},
  {Cl{\'e}ment}, {Ellis}, {Siana}, {Robertson}, {Schenker}, {Gutkin}, \&
  {Wofford}}]{Stark2014CIIIdetectionz67}
{Stark}, D.~P., {Richard}, J., {Charlot}, S., {et~al.} 2015{\natexlab{a}},
  \mnras, 450, 1846

\bibitem[{{Stark} {et~al.}(2015{\natexlab{b}}){Stark}, {Walth}, {Charlot},
  {Cl{\'e}ment}, {Feltre}, {Gutkin}, {Richard}, {Mainali}, {Robertson},
  {Siana}, {Tang}, \& {Schenker}}]{Stark2015CIV}
{Stark}, D.~P., {Walth}, G., {Charlot}, S., {et~al.} 2015{\natexlab{b}},
  \mnras, 454, 1393

\bibitem[{{Stark} {et~al.}(2017){Stark}, {Ellis}, {Charlot}, {Chevallard},
  {Tang}, {Belli}, {Zitrin}, {Mainali}, {Gutkin}, {Vidal-Garc{\'{\i}}a},
  {Bouwens}, \& {Oesch}}]{Stark2017}
{Stark}, D.~P., {Ellis}, R.~S., {Charlot}, S., {et~al.} 2017, \mnras, 464, 469

\bibitem[{{Stoughton} {et~al.}(2002){Stoughton}, {Lupton}, {Bernardi},
  {Blanton}, {Burles}, {Castander}, {Connolly}, {Eisenstein}, {Frieman},
  {Hennessy}, {Hindsley}, {Ivezi{\'c}}, {Kent}, {Kunszt}, {Lee}, {Meiksin},
  {Munn}, {Newberg}, {Nichol}, {Nicinski}, {Pier}, {Richards}, {Richmond},
  {Schlegel}, {Smith}, {Strauss}, {SubbaRao}, {Szalay}, {Thakar}, {Tucker},
  {Vanden Berk}, {Yanny}, {Adelman}, {Anderson}, {Anderson}, {Annis},
  {Bahcall}, {Bakken}, {Bartelmann}, {Bastian}, {Bauer}, {Berman},
  {B{\"o}hringer}, {Boroski}, {Bracker}, {Briegel}, {Briggs}, {Brinkmann},
  {Brunner}, {Carey}, {Carr}, {Chen}, {Christian}, {Colestock}, {Crocker},
  {Csabai}, {Czarapata}, {Dalcanton}, {Davidsen}, {Davis}, {Dehnen},
  {Dodelson}, {Doi}, {Dombeck}, {Donahue}, {Ellman}, {Elms}, {Evans}, {Eyer},
  {Fan}, {Federwitz}, {Friedman}, {Fukugita}, {Gal}, {Gillespie}, {Glazebrook},
  {Gray}, {Grebel}, {Greenawalt}, {Greene}, {Gunn}, {de Haas}, {Haiman},
  {Haldeman}, {Hall}, {Hamabe}, {Hansen}, {Harris}, {Harris}, {Harvanek},
  {Hawley}, {Hayes}, {Heckman}, {Helmi}, {Henden}, {Hogan}, {Hogg}, {Holmgren},
  {Holtzman}, {Huang}, {Hull}, {Ichikawa}, {Ichikawa}, {Johnston}, {Kauffmann},
  {Kim}, {Kimball}, {Kinney}, {Klaene}, {Kleinman}, {Klypin}, {Knapp},
  {Korienek}, {Krolik}, {Kron}, {Krzesi{\'n}ski}, {Lamb}, {Leger},
  {Limmongkol}, {Lindenmeyer}, {Long}, {Loomis}, {Loveday}, {MacKinnon},
  {Mannery}, {Mantsch}, {Margon}, {McGehee}, {McKay}, {McLean}, {Menou},
  {Merelli}, {Mo}, {Monet}, {Nakamura}, {Narayanan}, {Nash}, {Neilsen},
  {Newman}, {Nitta}, {Odenkirchen}, {Okada}, {Okamura}, {Ostriker}, {Owen},
  {Pauls}, {Peoples}, {Peterson}, {Petravick}, {Pope}, {Pordes}, {Postman},
  {Prosapio}, {Quinn}, {Rechenmacher}, {Rivetta}, {Rix}, {Rockosi}, {Rosner},
  {Ruthmansdorfer}, {Sandford}, {Schneider}, {Scranton}, {Sekiguchi}, {Sergey},
  {Sheth}, {Shimasaku}, {Smee}, {Snedden}, {Stebbins}, {Stubbs}, {Szapudi},
  {Szkody}, {Szokoly}, {Tabachnik}, {Tsvetanov}, {Uomoto}, {Vogeley}, {Voges},
  {Waddell}, {Walterbos}, {Wang}, {Watanabe}, {Weinberg}, {White}, {White},
  {Wilhite}, {Wolfe}, {Yasuda}, {York}, {Zehavi}, \& {Zheng}}]{Stoughton2022}
{Stoughton}, C., {Lupton}, R.~H., {Bernardi}, M., {et~al.} 2002, \aj, 123, 485

\bibitem[{{Sun} {et~al.}(2022){Sun}, {Egami}, {Fujimoto}, {Rawle}, {Bauer},
  {Kohno}, {Smail}, {P{\'e}rez-Gonz{\'a}lez}, {Ao}, {Chapman}, {Combes},
  {Dessauges-Zavadsky}, {Espada}, {Gonz{\'a}lez-L{\'o}pez}, {Koekemoer},
  {Kokorev}, {Lee}, {Morokuma-Matsui}, {Mu{\~n}oz Arancibia}, {Oguri},
  {Pell{\'o}}, {Ueda}, {Uematsu}, {Valentino}, {Van der Werf}, {Walth},
  {Zemcov}, \& {Zitrin}}]{Sun2022ApJ...932...77S}
{Sun}, F., {Egami}, E., {Fujimoto}, S., {et~al.} 2022, \apj, 932, 77

\bibitem[{{Thomas} {et~al.}(2005){Thomas}, {Maraston}, {Bender}, \& {Mendes de
  Oliveira}}]{Thomas2005}
{Thomas}, D., {Maraston}, C., {Bender}, R., \& {Mendes de Oliveira}, C. 2005,
  \apj, 621, 673

\bibitem[{{Trakhtenbrot} {et~al.}(2017){Trakhtenbrot}, {Volonteri}, \&
  {Natarajan}}]{Trakhtenbrot2017ApJ...836L...1T}
{Trakhtenbrot}, B., {Volonteri}, M., \& {Natarajan}, P. 2017, \apjl, 836, L1

\bibitem[{{Treu} {et~al.}(2015){Treu}, {Schmidt}, {Brammer}, {Vulcani}, {Wang},
  {Brada{\v{c}}}, {Dijkstra}, {Dressler}, {Fontana}, {Gavazzi}, {Henry},
  {Hoag}, {Huang}, {Jones}, {Kelly}, {Malkan}, {Mason}, {Pentericci},
  {Poggianti}, {Stiavelli}, {Trenti}, \& {von der Linden}}]{treu15}
{Treu}, T., {Schmidt}, K.~B., {Brammer}, G.~B., {et~al.} 2015, \apj, 812, 114

\bibitem[{{Treu} {et~al.}(2022){Treu}, {Roberts-Borsani}, {Bradac}, {Brammer},
  {Fontana}, {Henry}, {Mason}, {Morishita}, {Pentericci}, {Wang}, {Acebron},
  {Bagley}, {Bergamini}, {Belfiori}, {Bonchi}, {Boyett}, {Boutsia},
  {Calabr{\'o}}, {Caminha}, {Castellano}, {Dressler}, {Glazebrook}, {Grillo},
  {Jacobs}, {Jones}, {Kelly}, {Leethochawalit}, {Malkan}, {Marchesini},
  {Mascia}, {Mercurio}, {Merlin}, {Nanayakkara}, {Nonino}, {Paris},
  {Poggianti}, {Rosati}, {Santini}, {Scarlata}, {Shipley}, {Strait}, {Trenti},
  {Tubthong}, {Vanzella}, {Vulcani}, \& {Yang}}]{treu22}
{Treu}, T., {Roberts-Borsani}, G., {Bradac}, M., {et~al.} 2022, \apj, 935, 110

\bibitem[{{van der Walt} {et~al.}(2011){van der Walt}, {Colbert}, \&
  {Varoquaux}}]{vanderwalt11}
{van der Walt}, S., {Colbert}, S.~C., \& {Varoquaux}, G. 2011, Computing in
  Science Engineering, 13, 22

\bibitem[{{Vanden Berk} {et~al.}(2001){Vanden Berk}, {Richards}, {Bauer},
  {Strauss}, {Schneider}, {Heckman}, {York}, {Hall}, {Fan}, {Knapp},
  {Anderson}, {Annis}, {Bahcall}, {Bernardi}, {Briggs}, {Brinkmann}, {Brunner},
  {Burles}, {Carey}, {Castander}, {Connolly}, {Crocker}, {Csabai}, {Doi},
  {Finkbeiner}, {Friedman}, {Frieman}, {Fukugita}, {Gunn}, {Hennessy},
  {Ivezi{\'c}}, {Kent}, {Kunszt}, {Lamb}, {Leger}, {Long}, {Loveday}, {Lupton},
  {Meiksin}, {Merelli}, {Munn}, {Newberg}, {Newcomb}, {Nichol}, {Owen}, {Pier},
  {Pope}, {Rockosi}, {Schlegel}, {Siegmund}, {Smee}, {Snir}, {Stoughton},
  {Stubbs}, {SubbaRao}, {Szalay}, {Szokoly}, {Tremonti}, {Uomoto}, {Waddell},
  {Yanny}, \& {Zheng}}]{VandenBerk2001}
{Vanden Berk}, D.~E., {Richards}, G.~T., {Bauer}, A., {et~al.} 2001, \aj, 122,
  549

\bibitem[{{Vidal-Garc{\'\i}a} {et~al.}(2022){Vidal-Garc{\'\i}a}, {Plat},
  {Curtis-Lake}, {Feltre}, {Hirschmann}, {Chevallard}, \&
  {Charlot}}]{vidal-garcia22}
{Vidal-Garc{\'\i}a}, A., {Plat}, A., {Curtis-Lake}, E., {et~al.} 2022, arXiv
  e-prints, arXiv:2211.13648

\bibitem[{{Villa-V{\'e}lez} {et~al.}(2021){Villa-V{\'e}lez}, {Buat},
  {Theul{\'e}}, {Boquien}, \& {Burgarella}}]{Villa-Velez2021}
{Villa-V{\'e}lez}, J.~A., {Buat}, V., {Theul{\'e}}, P., {Boquien}, M., \&
  {Burgarella}, D. 2021, \aap, 654, A153

\bibitem[{{Virtanen} {et~al.}(2020){Virtanen}, {Gommers}, {Oliphant},
  {Haberland}, {Reddy}, {Cournapeau}, {Burovski}, {Peterson}, {Weckesser},
  {Bright}, {van der Walt}, {Brett}, {Wilson}, {Jarrod Millman}, {Mayorov},
  {Nelson}, {Jones}, {Kern}, {Larson}, {Carey}, {Polat}, {Feng}, {Moore}, {Vand
  erPlas}, {Laxalde}, {Perktold}, {Cimrman}, {Henriksen}, {Quintero}, {Harris},
  {Archibald}, {Ribeiro}, {Pedregosa}, {van Mulbregt}, \&
  {Contributors}}]{virtanen20}
{Virtanen}, P., {Gommers}, R., {Oliphant}, T.~E., {et~al.} 2020, Nature
  Methods, 17, 261

\bibitem[{{Voigt} \& {Bridle}(2010)}]{VoigtBridle2010MNRAS.404..458V}
{Voigt}, L.~M., \& {Bridle}, S.~L. 2010, \mnras, 404, 458

\bibitem[{{Volonteri}(2012)}]{Volonteri2012BHformation}
{Volonteri}, M. 2012, Science, 337, 544

\bibitem[{{Volonteri} {et~al.}(2023){Volonteri}, {Habouzit}, \&
  {Colpi}}]{Volonteri2022AGNz9}
{Volonteri}, M., {Habouzit}, M., \& {Colpi}, M. 2023, \mnras, 521, 241

\bibitem[{{Wang} {et~al.}(2023){Wang}, {Leja}, {Bezanson}, {Johnson},
  {Khullar}, {Labb{\'e}}, {Price}, {Weaver}, \& {Whitaker}}]{Wang2023}
{Wang}, B., {Leja}, J., {Bezanson}, R., {et~al.} 2023, \apjl, 944, L58

\bibitem[{{Wang} {et~al.}(2018){Wang}, {Yang}, {Fan}, {Yue}, {Wu}, {Schindler},
  {Bian}, {Li}, {Farina}, {Ba{\~n}ados}, {Davies}, {Decarli}, {Green}, {Jiang},
  {Hennawi}, {Huang}, {Mazzucchelli}, {McGreer}, {Venemans}, {Walter}, \&
  {Beletsky}}]{Wang2018z7Quasar}
{Wang}, F., {Yang}, J., {Fan}, X., {et~al.} 2018, \apjl, 869, L9

\bibitem[{Wang {et~al.}(2021)Wang, Yang, Fan, Hennawi, Barth, Banados, Bian,
  Boutsia, Connor, Davies, Decarli, Eilers, Farina, Green, Jiang, Li,
  Mazzucchelli, Nanni, Schindler, Venemans, Walter, Wu, \& Yue}]{Wang_2021}
Wang, F., Yang, J., Fan, X., {et~al.} 2021, The Astrophysical Journal Letters,
  907, L1.
\newblock \url{https://doi.org/10.3847%2F2041-8213%2Fabd8c6}

\bibitem[{{Weaver} {et~al.}(2023){Weaver}, {Cutler}, {Pan}, {Whitaker},
  {Labbe}, {Price}, {Bezanson}, {Brammer}, {Marchesini}, {Leja}, {Wang},
  {Furtak}, {Zitrin}, {Atek}, {Coe}, {Dayal}, {van Dokkum}, {Feldmann},
  {Forster Schreiber}, {Franx}, {Fujimoto}, {Fudamoto}, {Glazebrook}, {de
  Graaff}, {Greene}, {Juneau}, {Kassin}, {Kriek}, {Khullar}, {Maseda}, {Mowla},
  {Muzzin}, {Nanayakkara}, {Nelson}, {Oesch}, {Pacifici}, {Papovich}, {Setton},
  {Shapley}, {Smit}, {Stefanon}, {Taylor}, {Weibel}, \&
  {Williams}}]{Weaver2023}
{Weaver}, J.~R., {Cutler}, S.~E., {Pan}, R., {et~al.} 2023, arXiv e-prints,
  arXiv:2301.02671

\bibitem[{{Williams} {et~al.}(2023){Williams}, {Kelly}, {Chen}, {Brammer},
  {Zitrin}, {Treu}, {Scarlata}, {Koekemoer}, {Oguri}, {Lin}, {Diego}, {Nonino},
  {Hjorth}, {Langeroodi}, {Broadhurst}, {Rogers}, {Perez-Fournon}, {Foley},
  {Jha}, {Filippenko}, {Strolger}, {Pierel}, {Poidevin}, \&
  {Yang}}]{Williams2022z9p5}
{Williams}, H., {Kelly}, P.~L., {Chen}, W., {et~al.} 2023, Science, 380, 416

\bibitem[{{Windhorst} {et~al.}(2023){Windhorst}, {Cohen}, {Jansen}, {Summers},
  {Tompkins}, {Conselice}, {Driver}, {Yan}, {Coe}, {Frye}, {Grogin},
  {Koekemoer}, {Marshall}, {O'Brien}, {Pirzkal}, {Robotham}, {Ryan}, {Willmer},
  {Carleton}, {Diego}, {Keel}, {Porto}, {Redshaw}, {Scheller}, {Wilkins},
  {Willner}, {Zitrin}, {Adams}, {Austin}, {Arendt}, {Beacom}, {Bhatawdekar},
  {Bradley}, {Broadhurst}, {Cheng}, {Civano}, {Dai}, {Dole}, {D'Silva},
  {Duncan}, {Fazio}, {Ferrami}, {Ferreira}, {Finkelstein}, {Furtak}, {Gim},
  {Griffiths}, {Hammel}, {Harrington}, {Hathi}, {Holwerda}, {Honor}, {Huang},
  {Hyun}, {Im}, {Joshi}, {Kamieneski}, {Kelly}, {Larson}, {Li}, {Lim}, {Ma},
  {Maksym}, {Manzoni}, {Meena}, {Milam}, {Nonino}, {Pascale}, {Petric},
  {Pierel}, {del Carmen Polletta}, {R{\"o}ttgering}, {Rutkowski}, {Smail},
  {Straughn}, {Strolger}, {Swirbul}, {Trussler}, {Wang}, {Welch}, {B. Wyithe},
  {Yun}, {Zackrisson}, {Zhang}, \& {Zhao}}]{windhorst2022}
{Windhorst}, R.~A., {Cohen}, S.~H., {Jansen}, R.~A., {et~al.} 2023, \aj, 165,
  13

\bibitem[{{Yang} {et~al.}(2020){Yang}, {Boquien}, {Buat}, {Burgarella},
  {Ciesla}, {Duras}, {Stalevski}, {Brandt}, \& {Papovich}}]{Yang2020CIGALE1}
{Yang}, G., {Boquien}, M., {Buat}, V., {et~al.} 2020, \mnras, 491, 740

\bibitem[{{Yang} {et~al.}(2022){Yang}, {Boquien}, {Brandt}, {Buat},
  {Burgarella}, {Ciesla}, {Lehmer}, {Ma{\l}ek}, {Mountrichas}, {Papovich},
  {Pons}, {Stalevski}, {Theul{\'e}}, \& {Zhu}}]{Yang2022CIGALE2}
{Yang}, G., {Boquien}, M., {Brandt}, W.~N., {et~al.} 2022, \apj, 927, 192

\bibitem[{{Yang} {et~al.}(2023){Yang}, {Caputi}, {Papovich}, {Arrabal Haro},
  {Bagley}, {Behroozi}, {Bell}, {Bisigello}, {Buat}, {Burgarella}, {Cheng},
  {Cleri}, {Dave}, {Dickinson}, {Elbaz}, {Ferguson}, {Finkelstein}, {Grogin},
  {Hathi}, {Hirschmann}, {Holwerda}, {Huertas-Company}, {Hutchison}, {Iani},
  {Kartaltepe}, {Kirkpatrick}, {Kocevski}, {Koekemoer}, {Kokorev}, {Larson},
  {Lucas}, {Perez-Gonzalez}, {Rinaldi}, {Shen}, {Trump}, {de la Vega}, {Yung},
  \& {Zavala}}]{Yang2023CEERSAGN}
{Yang}, G., {Caputi}, K.~I., {Papovich}, C., {et~al.} 2023, arXiv e-prints,
  arXiv:2303.11736

\bibitem[{{Yang} {et~al.}(2021){Yang}, {Wang}, {Fan}, {Barth}, {Hennawi},
  {Nanni}, {Bian}, {Davies}, {Farina}, {Schindler}, {Ba{\~n}ados}, {Decarli},
  {Eilers}, {Green}, {Guo}, {Jiang}, {Li}, {Venemans}, {Walter}, {Wu}, \&
  {Yue}}]{Jinyi2021Quasars37z6}
{Yang}, J., {Wang}, F., {Fan}, X., {et~al.} 2021, \apj, 923, 262

\bibitem[{{Zackrisson} {et~al.}(2011){Zackrisson}, {Rydberg}, {Schaerer},
  {{\"O}stlin}, \& {Tuli}}]{Zackrisson2011PopIII}
{Zackrisson}, E., {Rydberg}, C.-E., {Schaerer}, D., {{\"O}stlin}, G., \&
  {Tuli}, M. 2011, \apj, 740, 13

\bibitem[{{Zheng} {et~al.}(2014){Zheng}, {Shu}, {Moustakas}, {Zitrin}, {Ford},
  {Huang}, {Broadhurst}, {Molino}, {Diego}, {Infante}, {Bauer}, {Kelson}, \&
  {Smit}}]{Zheng2014A2744}
{Zheng}, W., {Shu}, X., {Moustakas}, J., {et~al.} 2014, \apj, 795, 93

\bibitem[{{Zheng} {et~al.}(2017){Zheng}, {Wang}, {Rhoads}, {Infante},
  {Malhotra}, {Hu}, {Walker}, {Jiang}, {Jiang}, {Hibon}, {Gonzalez}, {Kong},
  {Zheng}, {Galaz}, \& {Barrientos}}]{LAGER2017ApJ...842L..22Z}
{Zheng}, Z.-Y., {Wang}, J., {Rhoads}, J., {et~al.} 2017, \apjl, 842, L22

\bibitem[{{Zitrin} {et~al.}(2014){Zitrin}, {Zheng}, {Broadhurst}, {Moustakas},
  {Lam}, {Shu}, {Huang}, {Diego}, {Ford}, {Lim}, {Bauer}, {Infante}, {Kelson},
  \& {Molino}}]{Zitrin2014highz}
{Zitrin}, A., {Zheng}, W., {Broadhurst}, T., {et~al.} 2014, \apjl, 793, L12

\bibitem[{{Zitrin} {et~al.}(2015){Zitrin}, {Labb{\'e}}, {Belli}, {Bouwens},
  {Ellis}, {Roberts-Borsani}, {Stark}, {Oesch}, \& {Smit}}]{Zitrin2015Lyalpha}
{Zitrin}, A., {Labb{\'e}}, I., {Belli}, S., {et~al.} 2015, \apjl, 810, L12

\end{thebibliography}

\end{document}